\documentclass[onecolumn,amsmath,amssymb,pra]{revtex4-1}

\usepackage{graphicx}
\usepackage{color}
\usepackage{dcolumn}
\usepackage{amsmath}
\usepackage{subfigure}

\usepackage{sidecap}
\usepackage{dsfont}

\usepackage{mathrsfs}
\usepackage{amsfonts}
\usepackage{indentfirst}
\usepackage{bm}

\usepackage[colorlinks,citecolor=blue]{hyperref}

\newcommand{\be}{\begin{equation}}
\newcommand{\ee}{\end{equation}}
\newcommand{\bey}{\begin{eqnarray}}
\newcommand{\eey}{\end{eqnarray}}
\newcommand{\bw}{\begin{widetext}}
\newcommand{\ew}{\end{widetext}}

\newcommand{\ra}{\rangle}
\newcommand{\la}{\langle}

\newcommand{\ba}{\begin{array}}
\newcommand{\ea}{\end{array}}
\newcommand{\bi}{\begin{itemize}}
\newcommand{\ei}{\end{itemize}}
\newcommand{\bem}{\begin{enumerate}}
\newcommand{\eem}{\end{enumerate}}

\begin{document}

\title{Characterizing dynamical phase transitions in a
spinor Bose-Einstein condensate via quantum and semiclassical analyses}

\author{Zhen-Xia Niu}
\affiliation{Department of Physics, Zhejiang Normal University, Jinhua 321004, China}

\author{Qian Wang}
\affiliation{Department of Physics, Zhejiang Normal University, Jinhua 321004, China, and \\
CAMTP-Center for Applied Mathematics and Theoretical Physics, University of Maribor, 
Mladinska 3, SI-2000 Maribor, Slovenia}

\begin{abstract}

Phase transitions in nonequilibrium dynamics of many body quantum systems, 
the so-called dynamical phases transition (DPTs), play an important role for understanding
various dynamical phenomena observed in different branches of physics.
In general, there have two types of DPTs, the first one refers to 
the phase transition that is characterized by 
distinct evolution behaviors of a physical observable, while the second one is marked by the 
nonanalyticities in the rate function of the initial state survival probability.   
Here, we focus on such DPTs from both quantum and semiclassical perspectives 
in a spinor Bose-Einstein condensate (BEC), an ideal platform to 
investigate nonequilibrium dynamics.
By using the sudden quench process, we demonstrate that the system exhibits 
both types of DPTs present as the control parameter quenches through
the critical one, referring to as the critical quench. 
We show analytically how to determine the critical quenches by means 
of the semiclassical approach 
and carry out a detailed examination on both semiclassical 
and quantum signatures of two types of DPTs. 
Moreover, we further reveal that the occurrence of DPTs is closely connected to
the separatrix in the underlying classical system. 
Our findings provide more insights into the properties of DPTs and verify the usefulness 
of semiclassical analysis for understanding DPTs in quantum systems 
with well-defined semiclassical limit.

\end{abstract}

\date{\today}

\maketitle

\section{Introduction}

The endeavor to reveal and understand exotic nonequilibrium phases 
of many body quantum systems 
has triggered a wide range of investigations of the dynamical phase transitions (DPTs). 
As an extension of the concept of equilibrium phase transition, DPTs refers 
the critical phenomena that happen in nonequilibrium dynamics of 
many body quantum systems \cite{Eckstein2008,Sciolla2011,Marino2022,Heyl2013,Heyl2018,Zvyagin2016}. 
The existence of DPTs has been verified in a variety of systems, 
such as the Hubbard model \cite{Sciolla2011,Moeckel2008,Eckstein2009,Sciolla2010}, 
the Dicke \cite{Lewis2021} and Rabi models \cite{Puebla2020}, 
different spin systems \cite{Zunkovic2016,Zunkovic2018,Sehrawat2021,Hashizume2022,
Karrasch2013,Schmitt2015,Zauner2017,Bhattacharya2017,Jafari2019,Mishra2020}, 
as well as Floquet systems \cite{Jafari2021,Jafari2022}, to name a few. 
Moreover, DPTs have also been observed in different experimental 
platforms involving trapped ions and cold atoms 
\cite{Jurcevic2017,ZhangJ2017,Scott2019,Muniz2020,TianT2020}.

A typical scenario for exploring of DPTs in an isolated system 
consists of the following steps. 
First, the system is prepared in an initial state, usually the ground state, 
at a certain value of a control parameter.
Then, the magnitude of the control parameter is suddenly changed 
from its initial value to a final value.
Depending on the quench strength, the resulting dynamics may exhibit 
two different kinds of critical behaviors,
dubbed as DPTs-I and DPTs-II, respectively. 
The first kind DPTs, denoted by DPTs-I, is signified by the changing of 
dynamics of a certain observable as the quench strength passes through the critical value 
\cite{Eckstein2008,Sciolla2011,Marino2022,Lewis2021,Puebla2020,
JFrank2018,Gambassi2011,Hailmeh2017,Lerose2019,
Corps2022,Corps2023a,Corps2023b}.
This leads to the using of long time average of such 
observable as the dynamical order parameter 
to define and distinguish different phases of DPTs-I, 
analogy to the equilibrium phase transitions.

Apart from DPTs-I, the dynamical critical behavior can also appear in the 
evolution of the initial state survival probability rate function, resulting in the concept 
of the second kind of DPTs \cite{Heyl2013,Heyl2018}, referred to as DPTs-II. 
The definition of DPTs-II is based on the mathematical equivalence between the amplitude of 
the initial state survival probability and the boundary partition function.   
As a consequence, the nonanalytical behavior in the rate function at particular
times is resembled to the singular behavior in the free energy at a critical temperature.
Hence, a DPT-II is characterized by nonanalyticities in the rate function 
at particular instants of times, called the critical times.
It is known that the occurrence of DPTs-II are independent of 
the equilibrium phase transitions \cite{Zunkovic2018,JafariR2019,Vajna2014,Canovi2014,Andraschko2014}, 
and shows a strong dependence on the initial condition \cite{Zauner2017,JFrank2018,Homrighausen2017}.
The connections between DPTs-I and DPTs-II have been discussed in several works
\cite{Zunkovic2018,Homrighausen2017,Weidinger2017,LangJ2018,
Corps2022,Corps2023a,Corps2023d},
however, a general relationship between them remains an open question.

In this work, we perform a detailed analysis of the properties of DPTs-I and DPTs-II 
in a spinor Bose-Einstein condensate (BEC)  \cite{Kawaguchi2012,Stamper2013}
from both quantum and semiclassical perspectives.
As a highly controllable platform, the spinor BECs have been extensively used, 
both theoretically and experimentally, to explore various nonequilibrium phenomena, 
such as the critical dynamics across the critical point of quantum phase transitions 
\cite{Damski2007,Lamacraft2007,Anquez2016,XueM2018,Bookjans2011,
KimJ2017,Prufer2018,ChenZ2019,LYQiu2020} 
and topological defects \cite{Sadler2006,Choi2012,KangS2019}.
Here, by using aforementioned standard protocol in the studies of DPTs 
to a ferromagnetic spin-$1$ BEC,  
we focus on the critical behaviors in the evolution of a certain observable, 
as well as in the initial state survival probability rate function.
We show that both kinds of DPTs are presented in the system 
as long as the sudden quench strength passes through the critical value. 
To uncover the triggering mechanism of DPTs, we analyze semiclassical limit of the system 
and discuss the correspondence between the quantum 
and semiclassical critical signatures of DPTs.
We further demonstrate that both kinds of DPTs are closely linked to
the separatrix in the underlying classical system. 
Hence, our results lead to a better understanding of the properties of DPTs and also  
verify the usefulness of the semiclassical analysis for studying DPTs 
in quantum systems with well-defined semiclassical limit.  
It is worth noting that the existence of DPT-I in antiferromagnetic spin-$1$ BEC has been explored 
from different aspects \cite{TianT2020,ZhouL2023,YangHX2019,HuangY2022}.
However, in this work, we are interested in both kinds of DPTs in ferromagnetic spin-$1$ BEC and 
show how to understand them from the semiclassical viewpoint.

In the following, after introducing basic features of the spin-$1$ BEC 
and reviewing its semiclassical limit in Sec.~\ref{second}, we report
our main results in Sec.~\ref{third}.
Specifically, we first present our employed quench protocol and explain how 
to obtain the critical quench strength using the semiclassical analysis.
Then, in Sec.~\ref{thirdA}, we provide a detailed investigation of the 
different signatures of DPT-I. 
We report our findings on DPT-II in Sec.~\ref{thirdB} 
and unveil the link between DPT-I and DPT-II.
Finally, we conclude in Sec.~\ref{fourth} with several remarks.

\section{Spin-1 spinor Bose condensate} \label{second}

The spinor Bose-Einstein condensates (BECs)
provide an excellent platform to explore a variety of interesting phenomena in condensate systems.
In this work, we focus on a spinor BEC consisting of $N$ spin-1 boses, such as $^{78}$Rb  
or $^{23}$Na atoms \cite{Chang2004,Stamper1998,Kawaguchi2012,Stamper2013}, in an optical trap. 
Under the single-mode approximation, which assumes that all spin 
states have a same spatial wave function\cite{YiS2002,Gabbrielli2015}, 
the system can be well described by the following Hamiltonian (setting $\hbar=1$) 
\cite{LawC1998,XueM2018,JieG2019}
\be \label{SBECH}
  \frac{H}{|c|}=\frac{\mathrm{sign}(c)}{N}\left[(a_1^\dag a_{-1}^\dag a_0^2+a_0^{\dag2}a_1 a_{-1})
          +N_0(N_{+1}+N_{-1})\right]+\xi(N_{-1}+N_{+1}).
\ee
Here, $a_m$ and $a_m^\dag$ are the bosonic annihilation and 
creation operators of state $m=0, \pm1$, 
$N_m=a_m^\dag a_m$ is the number operator and satisfies $\sum_m N_m=N$,
$c$ denotes the spin-dependent interaction strength with $c>0 (c<0)$ for 
the antiferromagnetic (ferromagnetic) atoms, and $\xi\equiv q/|c|$ 
is the rescaled quadratic Zeeman shift, 
which can vary between positive
and negative values via the microwave dressing \cite{Gerbier2006,Hamley2012}.

 \begin{figure}
  \includegraphics[width=\columnwidth]{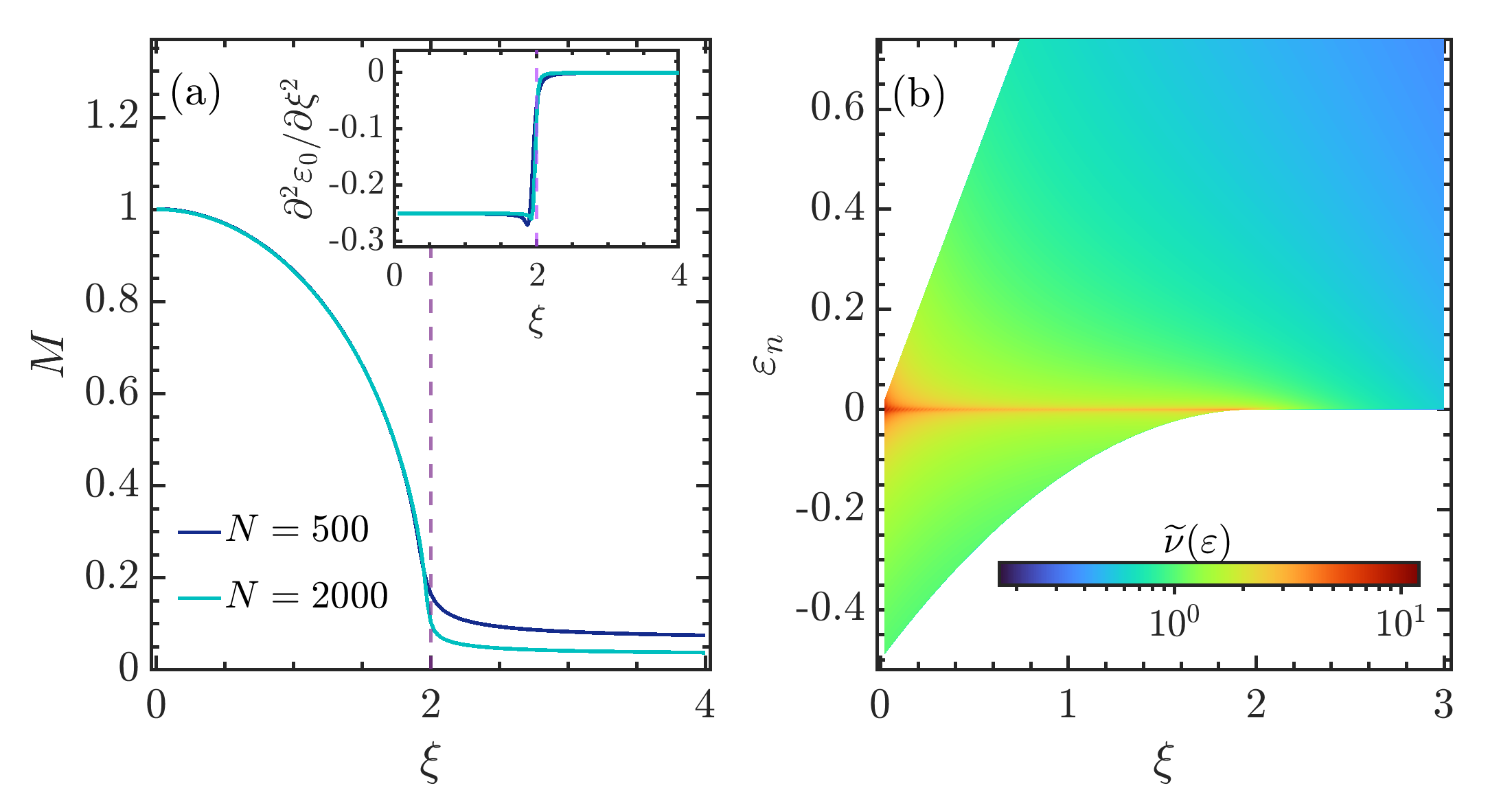}
  \caption{(a) Order parameter $M$ in (\ref{OrderP}) as a function of $\xi$ for different system sizes.
  Inset: $\partial^2\varepsilon_0/\partial\xi^2$ as a function of $\xi$ for 
  same system sizes as in main panel. Here, $\varepsilon_0=E_0/N$ 
  with $E_0$ denotes the ground state energy. 
  The vertical dashed lines in main panel and inset mark the critical value $\xi_c=2$.
  (b) Rescaled density of states 
  $\widetilde{\nu}(\varepsilon)=(1/N)\sum_n\delta(\varepsilon-\varepsilon_n)$ 
  as a function of $\xi$ and rescaled eigenenergy $\varepsilon_n=E_n/N$ for a system with $N=10000$.
  Here, the numerical result of $\widetilde{\nu}(\varepsilon)$ is obtained from Eq.~(\ref{CDOS}).  
  All quantities are dimensionless.
  }
  \label{PTs}
 \end{figure}

The Hamiltonian (\ref{SBECH}) conserves the total magnetization, 
$M=N_{+1}-N_{-1}$, and parity, $(-1)^{N_0}$.
Hence, we restrict our study in the subspace of $M=0$ with even parity, indicating that
the Hilbert space has dimension $\mathcal{D}_\mathcal{H}=N/2+1$ for even $N$. 
Moreover, we focus on the ferromagnetic condensate with $c<0$ and positive 
quadratic Zeeman shift $\xi\geq0$ in this work.

It is known that the ground state of Hamiltonian (\ref{SBECH}) undergoes a transition from the
broken-axisymmetry (BA) phase to the polar phase as $\xi$ passes 
through the critical point $\xi_c=2$
\cite{Stamper2013,XueM2018,Hamley2012,Sadler2006,ZhangZ2013,Feldmann2018}. 
This is signified by the behavior of the order parameter, defined as 
\cite{XueM2018,Damski2007,Lamacraft2007,Anquez2016}
\be \label{OrderP}
   M=\frac{\sqrt{\la F_x^2+F_y^2\ra}}{N},
\ee 
where $F_x=(F_++F_-)/2$, $F_y=(F_+-F_-)/(2i)$ with 
$F_+=\sqrt{2}(a_1^\dag a_0+a_0^\dag a_{-1})$ and $F_-=F_+^\dag$.
The evolution of $M$ as a function of $\xi$ for different system sizes is shown in Fig.~\ref{PTs}(a).
As an order parameter of the ground state quantum phase transition, 
$M$ is known for exhibiting a crossover from
$M\neq0$ in BA phase to $M=0$ in polar phase as we straddle the critical point. 
Further evidence of the quantum phase transition at $\xi_c=2$ is 
provided by the inset of Fig.~\ref{PTs}(a), where
we plot the second derivative of the rescaled ground state energy, 
$\varepsilon_0=E_0/N$, as a function of $\xi$.
An obvious sudden jump, which becomes more sharper with 
increasing $N$, around $\xi\approx2$ clearly marks 
the second order phase transition at $\xi_c=2$.

Apart from the ground state phase transition, the Hamiltonian (\ref{SBECH}) also exhibits 
an excited state quantum phase transition (ESQPT) in the BA phase 
\cite{Feldmann2021,NiuW2023,ZhouL2023,Meyer2023}. 
ESQPTs have been verified in a wide variety of many body 
quantum systems and show various impacts on 
both static and dynamical properties of systems, see Ref.~\cite{Cejnar2021} 
and references therein for a throughly review on different aspects of ESQPTs. 
First introduced as a generalization of ground state quantum phase transition, 
ESQPTs are characterized
by the singularity in the density of states (DOS) at the critical energy \cite{Caprio2008}.
For our studied spinor BEC, this is confirmed in Fig.~\ref{PTs}(b), where we plot how the
rescaled DOS, $\widetilde{\nu}(\varepsilon)=(1/N)\sum_n\delta(\varepsilon-\varepsilon_n)$, 
varies as a function of $\xi$ and rescaled energy $\varepsilon_n=E_n/N$. 
One see an obvious peak around $\varepsilon=0$ in the behavior of 
$\widetilde{\nu}(\varepsilon)$ as long as $\xi<\xi_c=2$. 
The peak observed in $\widetilde{\nu}(\varepsilon)$ serves 
as a finite size precursors of ESQPT and turns into
the logarithmic divergence, $\widetilde{\nu}(\varepsilon)\propto-\ln|\varepsilon-\varepsilon_c|$ 
with $\varepsilon_c=0$,
in the thermodynamic limit $N\to\infty$ \cite{Feldmann2021,Caprio2008,Stransky2014}.

 \begin{figure}
  \includegraphics[width=\columnwidth]{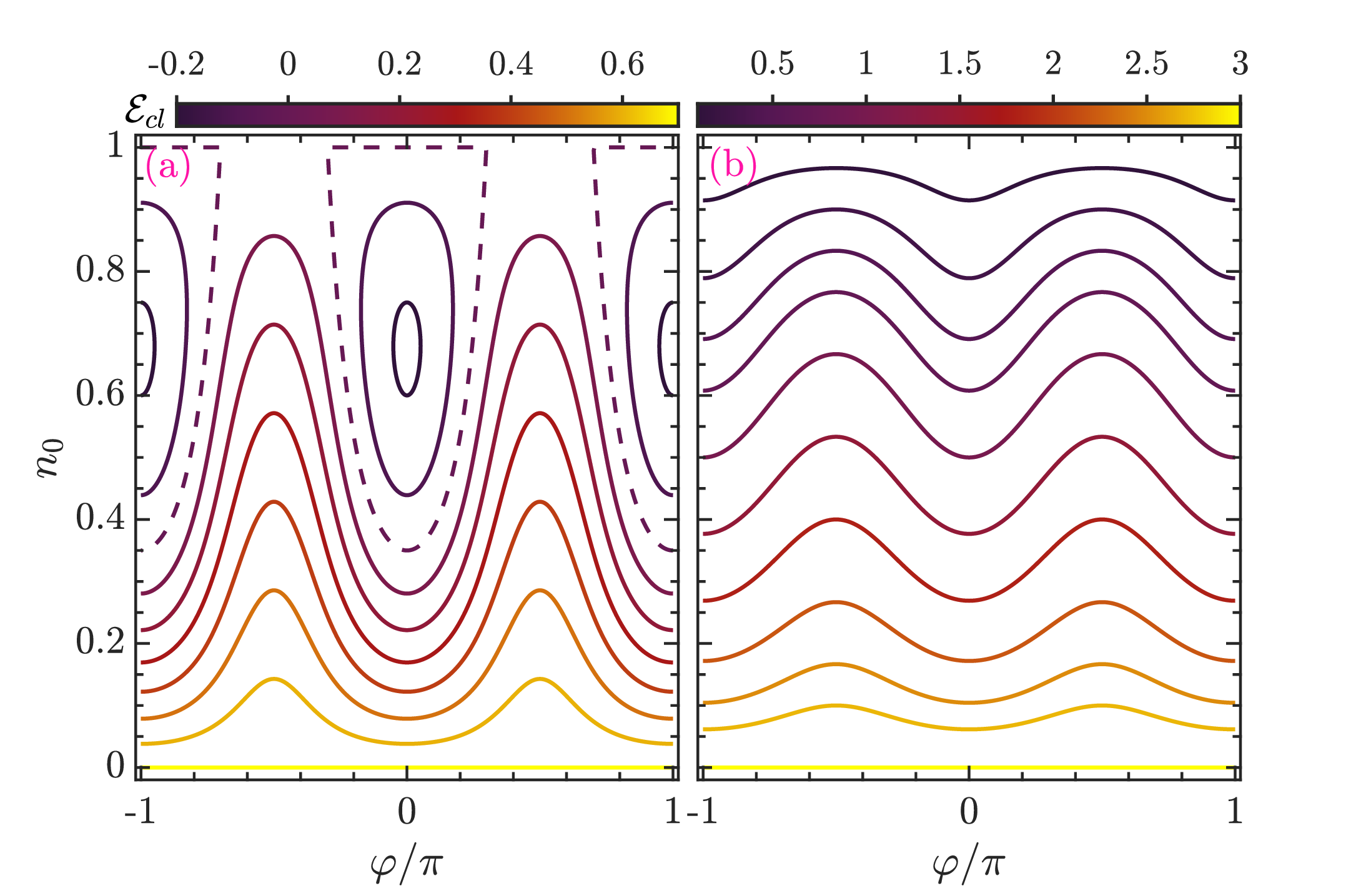}
  \caption{Contour plot in the phase space of the classical Hamiltonian (\ref{ClassicalH}) for
  (a) $\xi=1$ and (b) $\xi=3$.
  The energy for each contour is indicated by the color bar.
  The dashed line in panel (a) marks the separatrix determined by $\mathcal{E}_{cl}=0$. 
  Panel (a) illustrates that the classical system has two degenerate 
  minima and exhibits a separatrix in its dynamics.
  Panel(b) shows a connected phase space with minima is unique for the case of $\xi>\xi_c$. 
  All quantities are dimensionless.       
  }
  \label{CEg}
 \end{figure}

\subsection*{Semiclassical limit}

More insights into the properties of the system and associated 
phase transitions can be gained from the semiclassical analysis in the limit $N\to\infty$. 
To this end, we calculate the semiclassical counterpart of the Hamiltonian (\ref{SBECH}) 
by using the coherent states, defined as
\cite{ZhangW2005,Rautenberg2020,Feldmann2021,ZhouL2023,Yulong2021}
\begin{align} \label{Coherents}
  |\bm{\alpha}\ra&=\frac{1}{\sqrt{N!}}\left(\sum_m\alpha_m a_m^\dag\right)^N|0\ra,
\end{align}
where $|0\ra$ is the bosonic vacuum state, so that $a_m|0\ra=0$.
Here, $\alpha_m=\sqrt{n_m}e^{i\phi_m}$ with $\phi_m\in[-\pi,\pi)$, 
$n_m\in[0,1]$ and satisfies $\sum_m n_m=1$. 
For our considered case with $N_{-1}=N_1$, the coherent state 
in Eq.~(\ref{Coherents}) can be explicitly expressed as
\be
    |\bm{\alpha}\ra=\sum_{N_0=0}^N\binom{N}{N_0}^{1/2}(n_0)^{\frac{N_0}{2}}
             (1-n_0)^{\frac{N-N_0}{2}}e^{-iN_0\varphi}|N,N_0\ra,
\ee
with $N_{-1}=N_1=(N-N_0)/2$ and $|N,N_0\ra$ are the Fock states. 
The classical limit of the model is given by the expectation value of $H$ in Eq.~(\ref{SBECH}) 
with respect to the coherent state in the classical limit $N\to\infty$. 
By employing the relation $\la\bm{\alpha}|a_m^\dag a_{m'}|\bm{\alpha}\ra=N\alpha_m^\ast\alpha_{m'}$
\cite{Yulong2021,Feldmann2021}, one can easily find that the classical Hamiltonian 
with $N_1=N_{-1} (n_{-1}=n_1)$ reads \cite{ZhangW2005,Feldmann2021}
\be \label{ClassicalH}
   \mathcal{H}_{cl}=\frac{1}{|c|}\lim_{N\to\infty}\frac{\la\bm{\alpha}|H|\bm{\alpha}\ra}{N}
                         =\xi(1-n_0)-2n_0(1-n_0)\cos^2\varphi,
\ee
where $\varphi=\phi_0-(\phi_1+\phi_{-1})/2$ with $\phi_1=\phi_{-1}$ is the relative phase.
The classical equations of motion are
\begin{align}\label{EOM}
  \dot{\varphi}&=\frac{\partial\mathcal{H}_{cl}}{\partial n_0}=-2(1-2n_0)\cos^2\varphi-\xi, \notag \\
  \dot{n}_0&=-\frac{\partial\mathcal{H}_{cl}}{\partial\varphi}=-2n_0(1-n_0)\sin(2\varphi),
\end{align}
associated with constrained condition $d(\phi_1-\phi_{-1})/dt=0$.

 \begin{figure}
  \includegraphics[width=\columnwidth]{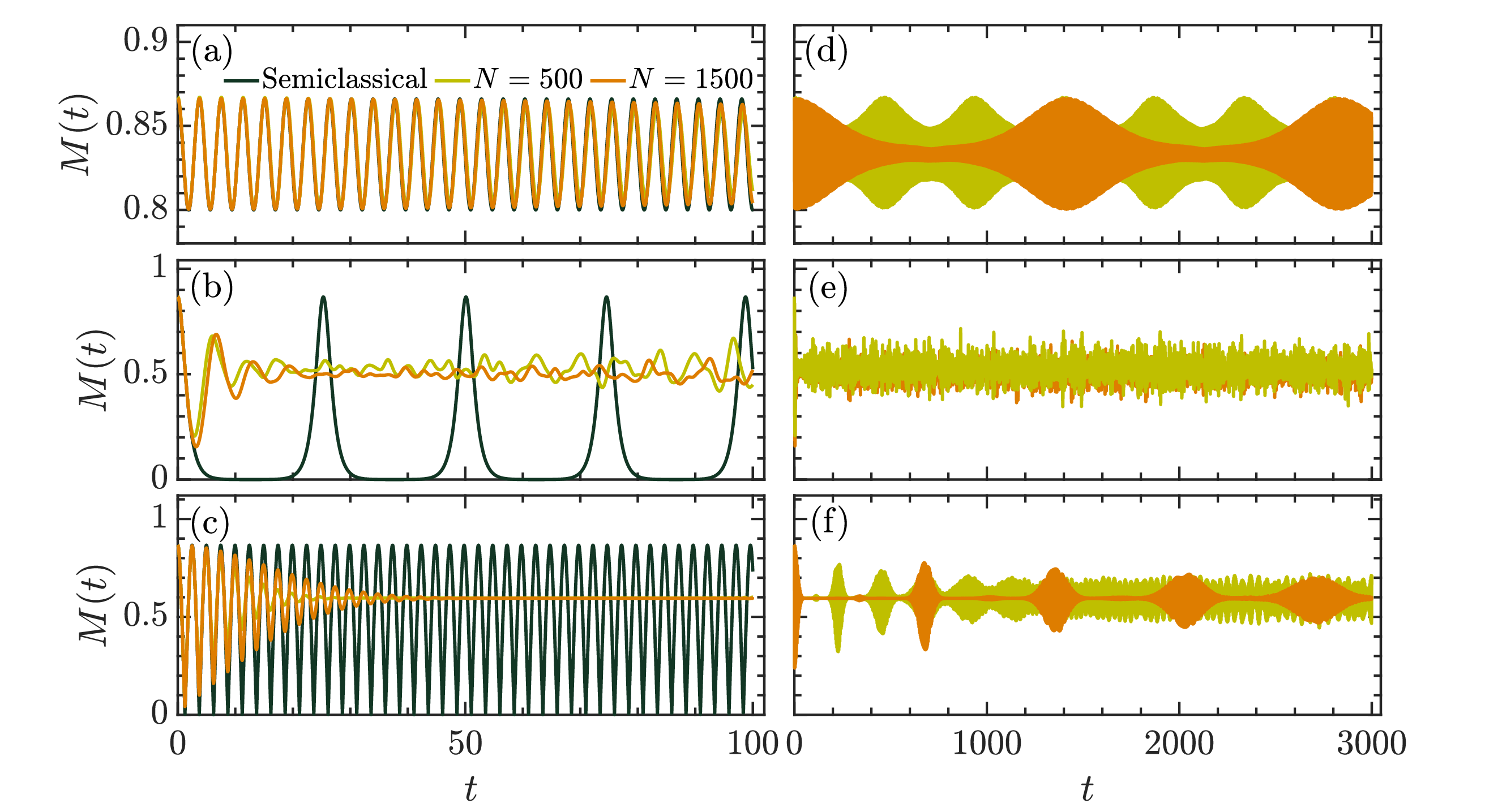}
  \caption{Dynamics of the observable $M(t)$ (\ref{OrderP}) 
  for different system sizes with [(a)(d)] $\delta\xi=0.1$, 
  [(b)(e)] $\delta\xi=0.5$, and [(c)(f)] $\delta\xi=1$.  
  The color code in panel (a) is followed in all panels. 
  In panels (a)-(c), dark green lines denote the semiclassical results. 
  The initial state is the ground state of the system (\ref{SBECH}) with $\xi=\xi_i=1$ 
  and the critical quench strength is $\delta\xi_c=0.5$ [see Eq.~(\ref{CriticalXi})].
  The axes in all figures are dimensionless.     
  }
  \label{TimeMt}
 \end{figure}

The fixed points of the classical system are the values $(\varphi_s,n_{0,s})$
that lead to $\dot{\varphi}=\dot{n}_0=0$.
We first note that two fixed points $n_{0,s}=0,1$ are independent of $\xi$ and $\varphi$ values.
The associated classical energy of them are given by $\mathcal{E}_{cl}=\xi$ and $0$, respectively.
However, the stability analysis tells us that the point $n_{0,s}=0$ is an unstable fixed point for $\xi>0$. 
On the contrary, the point $n_{0,s}=1$ is stabled for $\xi\geq\xi_c=2$, 
while it becomes unstable when $\xi<\xi_c$.
For the case of $\xi<\xi_c$, the stable fixed point of the system is provided 
by two degenerate points $(\cos\varphi_s,n_{0,s})=[\pm1,(2+\xi)/4]$, 
corresponding to classical energy $\mathcal{E}_{cl}=-(2-\xi)^2/8$.
As a consequence, the rescaled ground state energy, which 
corresponds to the classical energy evaluated at the
stable fixed points, can be written as
\begin{align}
  \varepsilon_0=\mathcal{E}_{cl,s}=
  \left\{
  \begin{aligned}
   &-\frac{(2-\xi)^2}{8}\quad \text{for}\ \xi<\xi_c, \\
   &0\quad \text{for}\ \xi\geq\xi_c.
  \end{aligned}
  \right.
\end{align}
Clearly, the second derivation of $\varepsilon_0$ with respect to $\xi$ 
undergoes a jump at $\xi_c=2$, in agreement with the numerical results
shown in the inset of Fig.~\ref{PTs}(a). 
Moreover, in the classical limit with $N\to\infty$ the order parameter $M$ reads \cite{Hoang2016} 
\begin{align} \label{ClassicalOP}
  \mathcal{M}_{cl}=\sqrt{4n_0(1-n_0)\cos^2\varphi}=
  \left\{
  \begin{aligned}
  &\frac{1}{2}\sqrt{4-\xi^2}\quad\ \text{for}\ \xi<\xi_c,  \\
  &0,\quad \text{for}\ \xi\geq\xi_c.
  \end{aligned}
  \right.
\end{align}
It consists with the observed behaviors of $M$ in the main panel of Fig.~\ref{PTs}(a).

In Fig.~\ref{CEg}, we show the energy contours in the phase space of the classical system (\ref{ClassicalH})
for different values of control parameter $\xi$.
Each curve stands for the set of points $(\varphi,n_0)$ that satisfies 
$\mathcal{H}_{cl}(\varphi,n_0)=\varepsilon$ with $\varepsilon$ is a given energy. 
Note that the phase space for $\xi>\xi_c$ case is featureless and exhibits a 
global minimum at $n_0=1$, as seen in Fig.~\ref{CEg}(b).
However, as $n_0=1$ becomes a saddle point and two degenerate minima points appear for
$\xi<\xi_c$, the phase space in this case exhibits a complex structure,
as demonstrated in Fig.~\ref{CEg}(a). 
The emergence of saddle point at $n_0=1$ with $\mathcal{E}_{cl}=0$ 
implies the presence of separatrix in the underlying classical dynamics, 
as marked by the dashed lines in Fig.~\ref{CEg}(a).

 \begin{figure}
  \includegraphics[width=\columnwidth]{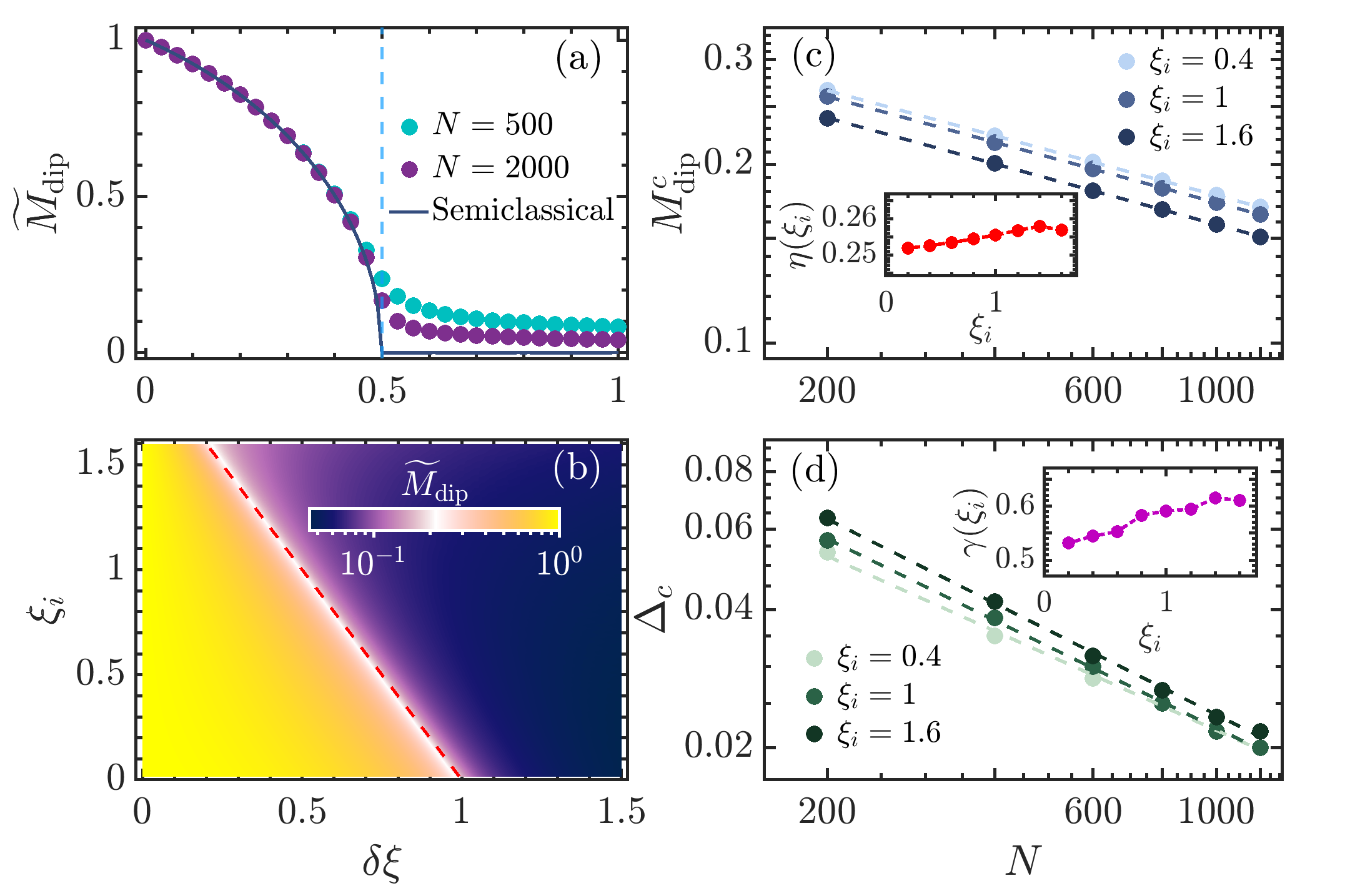}
  \caption{(a) Rescaled $M_\mathrm{dip}$, 
  $\widetilde{M}_\mathrm{dip}=M_\mathrm{dip}/M_\mathrm{dip}^m$,
  as a function of $\delta\xi$ for different system sizes with $\xi_i=1$, 
  which yields $\delta\xi_c=0.5$ [see Eq.~(\ref{CriticalXi})].
  Here, $M_\mathrm{dip}^m$ is the maximal value of $M_\mathrm{dip}$.
  The solid line denotes the semiclassical result and the vertical dashed line marks $\delta\xi_c=0.5$.
  (b) $\widetilde{M}_\mathrm{dip}$ as a function of $\delta\xi$ and $\xi_i$ with $N=1000$.
  The red dashed line marks the analytical critical line in (\ref{CriticalXi}).
  (c) Finite system size value of $M_\mathrm{dip}$ at $\delta\xi_c$, 
  denoted by $M_\mathrm{dip}^c$, as a function of
  $N$ for several $\xi_i$ values. The dashed lines represent the power low behavior of the form
  $M_\mathrm{dip}^c\propto N^{-\eta(\xi_i)}$ with the dependence of the exponents 
  $\eta(\xi_i)$ on $\xi_i$ is shown in the inset. 
  (d) Scaling of $\Delta_c=|\delta\xi_c(N)-\delta\xi_c|$ with 
  system size $N$ for several values of $\xi_i$.
  Here, $\delta\xi_c(N)$ is the precursor of the critical point and identified as the location of 
  the minimal value of $\partial M_\mathrm{dip}/\partial(\delta\xi)$.
  The dashed lines are the power law fitting, $\Delta_c\propto N^{-\gamma(\xi_i)}$. 
  The inset plots how $\gamma(\xi_i)$ varies as a function of $\xi_i$. 
  All quantities are dimensionless.   
  }
  \label{MdipPD}
 \end{figure}

The coincidence between the critical energy of ESQPT and separatrix indicates that
the occurrence of ESQPT can attribute to the existence of the saddle point in 
the corresponding classical system \cite{Caprio2008,Stransky2014}.    
This can be verified by the available phase space volume $\nu_{cl}(\varepsilon)$, 
which is the classical approximation 
of the quantum density of states \cite{Gutzwiller2013} and for our considered system is defined as
\cite{Feldmann2021}
\be\label{CDOS}
    \nu_{cl}(\varepsilon)=\frac{1}{2\pi}\int dn_0d\varphi
       \delta(\varepsilon-\mathcal{H}_{cl}).
\ee
Here, we have taken into account the condition $n_{-1}=n_1$.  
One can evaluate the above integral by using the properties of the Dirac delta function. 
The leading behavior of $\rho_{cl}(\varepsilon)$ around the ESQPT 
critical energy $\varepsilon_c=0$ 
is given by \cite{Kawaguchi2012,ZhangW2005,Ribeiro2008}
\be
  \lim_{\varepsilon\to\varepsilon_c}\nu_{cl}(\varepsilon)=
  -\frac{\ln|\varepsilon-\varepsilon_c|}{2\pi\sqrt{\xi(2-\xi)}}.
\ee
This asymptotic behavior confirms the close connection between the 
ESQPT in quantum system and
the saddle point in the underlying classical system.

Above discussion demonstrates the usefulness of semiclassical approach for studying  
the characterizations of phase transitions in quantum model and inspires us to
investigate whether it can help us to get a better understanding of the 
signatures of dynamical phase transitions (DPTs). 
In the rest of this article, we perform a detailed exploration on the DPTs in spinor BEC
and show how to understand the main characters of DPTs from the corresponded classical dynamics.

\section{Dynamical phase transitions} \label{third}

The term DPTs refer two kinds of phase transitions, 
denoted by DPTs-I and DPTs-II, respectively.  
The first kind, DPTs-I, is identified by the non-equilibrium order parameter, which defines 
as a long-time average of a certain observable and undergoes 
an abrupt change at the critical point that divides different dynamical phases
\cite{Eckstein2009,Sciolla2010,Gambassi2011,Sciolla2011,Sciolla2013,
Hailmeh2017,Lerose2019,Puebla2020,
Lewis2021,Corps2022,Corps2023a,Corps2023b}.  
Typically, DPTs-I are studied by means of a quantum quench, which is a sudden change of control parameters in the system
and results in the non-equilibrium dynamics. 
It is closely related to prethermalization \cite{Gring2012,Mori2018} 
and has been observed in numerous experiments
\cite{ZhangJ2017,Scott2019,Muniz2020,TianT2020,Marino2022}. 

Despite the second type, DPTs-II, also induced by the sudden quench 
process in isolated quantum systems,
it is characterized by singularities in the Loschmidt echo 
rate function at critical times \cite{Heyl2013,Heyl2018}, 
instead of the nonanalytical behaviors in the dynamical order parameter.  
It was originally defined in the one-dimensional Ising model and 
attracted a great deal of effort to study various aspects
in a variety of quantum systems \cite{Corps2022,Corps2023a,Corps2023b,Karrasch2013,Heyl2014,Schmitt2015,Zauner2017,
Bhattacharya2017,Jurcevic2017,Jafari2019,Mishra2020,Nicola2021,Jafari2021,Jafari2022,Corps2023c}.
The DPTs are independent of the equilibrium phase transitions \cite{Sciolla2011,Vajna2014},
but a general relationship between two types of DPTs is still an open question
\cite{Lerose2019,Zunkovic2016,Zunkovic2018,LangJ2018,Sehrawat2021,Hashizume2022}.

In this work, we focus on both types of DPTs in the spin-1 spinor BEC.
To this end, we consider the following quench process.
The system is initially prepared in the ground state, $|\psi_0\ra$, of the initial Hamiltonian $H_i$ obtained by
fixing $\xi=\xi_i<\xi_c$ in (\ref{SBECH}). 
Then, at $t=0$, we suddenly change the value of $\xi$ from its initial value $\xi_i$ 
to a certain final value $\xi_f=\xi_i+\delta\xi$. 
Now, as the initial state is not the eigenstate of the final Hamiltonian 
$H_f=H(\xi_f)$ anymore, it starts to evolved.
At time $t$, the density of state of the system is given by 
\be\label{densityt}
   \rho(t)=|\psi_t\ra\la\psi_t|=e^{-iH_ft}|\psi_0\ra=\sum_{n,k}c_n^\ast c_{k}
   e^{-i(E_{k}^f-E_n^f)t}|E_k^f\ra\la E_n^f|,
\ee
where $|E_k^f\ra$ is the $k$th eigenstate of $H_f$ with energy $E_k^f$, so that 
$H_f|E_k^f\ra=E_k^f|E_k^f\ra$, and $c_k=\la E_k^f|\psi_0\ra$ 
is the overlap between $|E_k^f\ra$ and the initial state.

 \begin{figure}
  \includegraphics[width=\columnwidth]{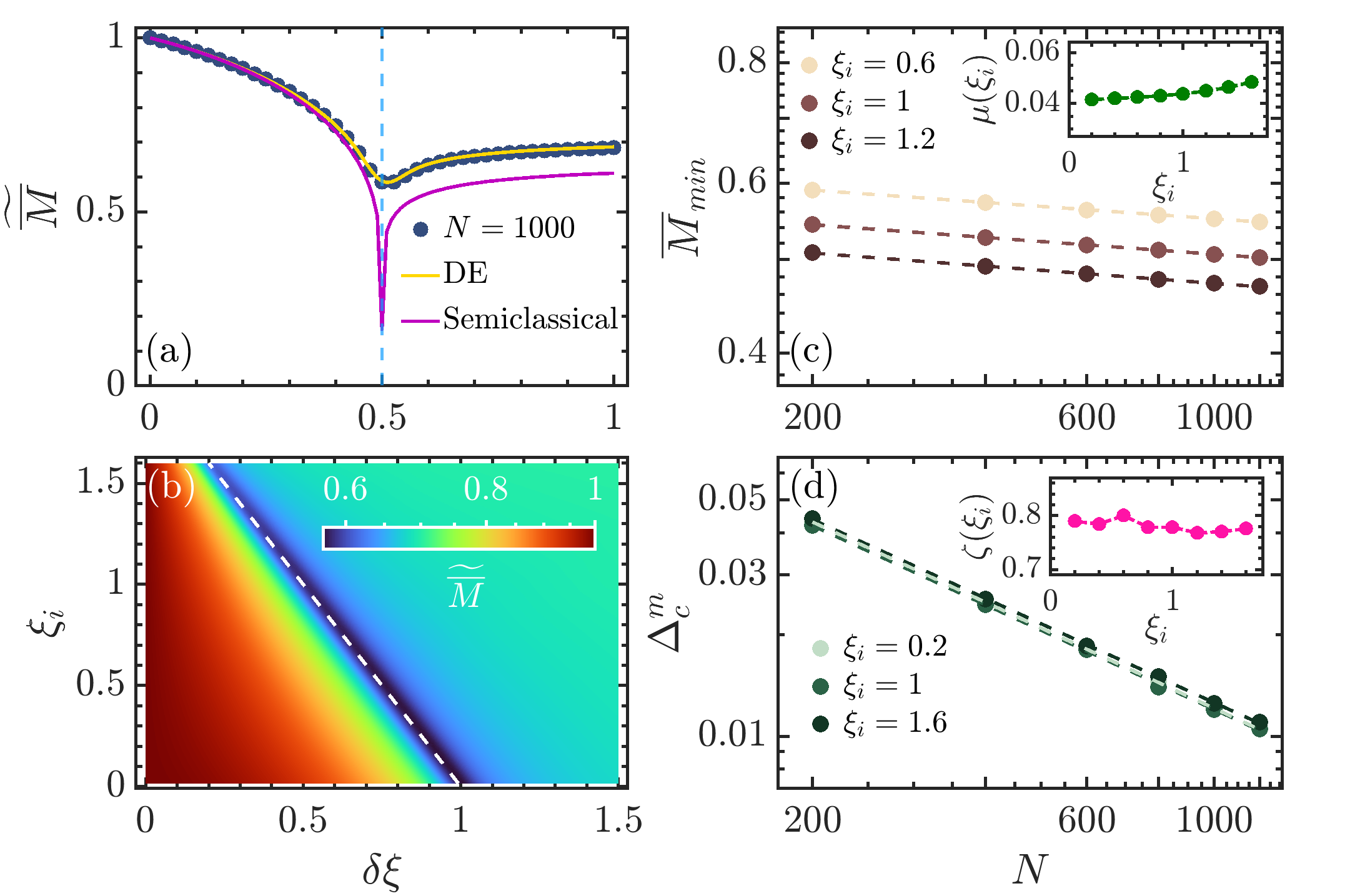}
  \caption{(a) Rescaled long time average of $M(t)$ in (\ref{AvgM}), 
  $\widetilde{\overline{M}}=\overline{M}/\overline{M}_{max}$,
  as a function of $\delta\xi$ for $\xi_i=1$, so that $\delta\xi_c=0.5$ 
  [see Eq.~(\ref{CriticalXi})], marked by vertical dashed line.
  Here, $\overline{M}_{max}$ denotes the maximal value of $\overline{M}$.
  Full circles are the numerical results for $N=1000$, obtained in the time interval $t\in[0,1000]$. 
  The yellow solid line is obtained using the diagonal ensemble average in (\ref{DEavg}), 
  while the purple solid line represents the semiclassical result.
  (b) $\widetilde{\overline{M}}$ as a function of $\xi_i$ and $\delta\xi$ with $N=1000$.
  The white dashed line marks the analytical critical line in Eq.~(\ref{CriticalXi}).   
  (c) Minimal value of $\overline{M}$, denoted by $\overline{M}_{min}$, as a function of
  $N$ for several $\xi_i$ values. The dashed lines represent the power low scaling of the form
  $\overline{M}_{min}\propto N^{-\mu(\xi_i)}$ with the 
  dependence of $\mu(\xi_i)$ on $\xi_i$ is shown in the inset. 
  (d) Scaling of $\Delta^m_c=|\delta\xi_c^m(N)-\delta\xi_c|$ with 
  system size $N$ for several values of $\xi_i$.
  Here, $\delta\xi_c^m(N)$ is the precursor of the critical point and 
  it has been identified as the position of $\overline{M}_{min}$.
  The dashed lines are the power law fitting, $\Delta_c^m\propto N^{-\zeta(\xi_i)}$. 
  The variation of $\zeta(\xi_i)$ with $\xi_i$ is plotted in the inset. 
  The axes in all figures are dimensionless. 
  }
  \label{PDavM}
 \end{figure}

The sudden quench of the value of control parameter from $\xi_i$ to $\xi_f$ leads to the change of the initial state energy. 
The energy of the quenched initial state can be calculated using the classical approach with the result given by 
\be
   \mathcal{E}_f(\xi_i,\delta\xi)=\frac{1}{N}\la\psi_0|H_f|\psi_0\ra=\mathcal{H}_{cl}^f|_{(\varphi_s,n_{0,s})},
\ee
where $\mathcal{H}_{cl}^f$ is the classical limit version of $H_f$ and $(\varphi_s,n_{0,s})$ 
is the initial state corresponded fixed point.
Then, by inserting $\varphi_s=\arccos(\pm1)$ and $n_{0,s}=(\xi_i+2)/4$ into above equation, 
it is straightforward to find that the explicit form of $\mathcal{E}_f(\xi_i,\delta\xi)$ can be expressed as
\be
  \mathcal{E}_f(\xi_i,\delta\xi)=\frac{2-\xi_i}{8}\left[2\delta\xi-(2-\xi_i)\right].
\ee  
Hence, the energy of the quenched system can be varied by tuning the quench strength $\delta\xi$.
The existence of the separatrix for $\xi_i<2$ in the classical dynamics implies that there has a critical 
quench strength $\delta\xi_c$, which
takes the quenched system to the energy of the separatrix, which is given by $\mathcal{E}_{cl,sp}=0$. 
For the energy below $\mathcal{E}_s$ the classical phase space possesses two disjointed regions, while these regions
are merged together for $\mathcal{E}_f(\xi_i,\delta\xi)>\mathcal{E}_{cl,sp}$, see Appendix~\ref{Appa} for further details.
From the condition $\mathcal{E}_f(\xi_i,\delta\xi)=\mathcal{E}_{cl,sp}=0$, 
one can easily find that the critical quench strength is given by
\be \label{CriticalXi}
   \delta\xi_c=1-\frac{\xi_i}{2},
\ee
with $0<\xi_i<2$.
In the following of this section, we investigate the consequences of 
the separatrix in nonequilibrium dynamics with the aim to explore various dynamical phases.

\subsection{DPTs-I: dynamical order parameters} \label{thirdA}

The first type of DPTs, DPTs-I, is signified by different evolution behavior of 
a physically relevant observable as the system control parameter varies. 
Here, we choose $M$ in Eq.~(\ref{OrderP}) as our studied observable.
The discussion for the case of other observable can be found in Appendix \ref{Appb}.
After quench, the time evolution of $M$ reads
\be\label{MtEv}
  M(t)=\mathrm{Tr}[\rho(t)M]
    =\sum_{n,k}c_n^\ast c_ke^{-i(E^f_k-E^f_n)t}M_{kn},
\ee
where $M_{nk}=\la E_k^f|M|E_n^f\ra$.
We also study the evolution of $\mathcal{M}_{cl}(t)$ in (\ref{ClassicalOP}), which is the classical counterpart of $M(t)$
and is completely governed by the classical equations of motion Eq.~(\ref{EOM}).
As the quantum dynamics approaches the classical one with increasing the system size $N$, 
the evolution of $M(t)$ should be well captured by $\mathcal{M}_{cl}(t)$ in the large $N$ limit.

In Figs.~\ref{TimeMt}(a)-\ref{TimeMt}(c), we plot the time evolutions of $M(t)$ 
for several quenching strengths and system sizes with $\xi_i=1$, 
which yields $\delta\xi_c=0.5$ according to Eq.~(\ref{CriticalXi}).
The evolutions of $\mathcal{M}_{cl}(t)$ for each case are depicted with dark green lines.   
Overall, the behavior of $M(t)$ exhibits a strong dependence on the quenching strength.
Specifically, for $\delta\xi<\delta\xi_c=0.5$, as illustrated in 
Fig.~\ref{TimeMt}(a), $M(t)$ undergoes a regular oscillation
with small amplitude and follows the classical dynamics even for relatively small system size.
This due to the fact that the quenched state for small quenching strengths 
remains within one of the two classical wells
and oscillates around the initial fixed point with a same frequency as the classical counterpart.     
On the contrary, the evolution of $M(t)$ undergoes a remarkable change for the 
quenching strengths that above the critical value, such as 
$\delta\xi=1$ case plotted in Fig.~\ref{TimeMt}(c).  
For quenches with $\delta\xi>\delta\xi_c$, the quenched 
system has enough energy and the corresponded 
classical dynamics can explore both wells in the phase space.
As a result, the evolution of $\mathcal{M}_{cl}(t)$ oscillates regularly with large amplitude. 
However, we see damping combined with dephasing in the quantum dynamics. 
Moreover, the deviation time between the quantum and classical dynamics is very short, 
even for large system size.
The distinct difference in the evoliutions of $M(t)$ and $\mathcal{M}_{cl}(t)$ suggests that
quantum correlations have significant impacts on 
the quantum dynamics when $\delta\xi>\delta\xi_c$.  
The critical quench is demonstrated in Fig.~\ref{TimeMt}(b), 
where the quenched system has energy coincides with
the energy of separatrix.
Classically, the evolution of the system will spend long time 
at the saddle point $\rho_{0,f}=1$, leading to
the low frequency of the oscillation of $\mathcal{M}_{cl}(t)$.
A drastic deviation from the behavior of $\mathcal{M}_{cl}(t)$ 
in the evolution of $M(t)$ can be clearly identified in this critical case.
The separatrix in the underlying classical system leads to a quite fast equilibration 
process in the quantum dynamics  

The longer time evolution of the same state as in Figs.~\ref{TimeMt}(a)-\ref{TimeMt}(c) 
are shown in
Figs.~\ref{TimeMt}(d)-\ref{TimeMt}(f).
As seen in the short time behaviors, the evolution of $M(t)$ in the 
longer time scale still depends on
whether the value of $\delta\xi$ is below, at, or above the critical value.  
Generally, for noncritical quenches, $M(t)$ oscillates around its steady state value 
and shows dynamical revivals at later times, as clearly 
seen from Figs.~\ref{TimeMt}(d) and \ref{TimeMt}(f).
The dynamical revival is an echo of the short time behaviors of a physical quantity and 
is a consequence of the discreteness in the spectrum of the system \cite{Milburn1997,Veksler2015}.
The time for the first revival and the time interval between two succeeding revivals 
increase as the system size $N$ increases.  
One can expect that the revival patterns will disappear and 
$M(t)$ behaves very noisy at very long times.
This is already visible in Fig.~\ref{TimeMt}(f) for $N=500$ case.
The situation for the critical quench in Fig.~\ref{TimeMt}(e) shows a complete different scenario.
There is no revival in quantum dynamics and $M(t)$ evolves with 
erratic fluctuations around its equilibrium value. 

The results displayed in Fig.~\ref{TimeMt} neatly manifest the 
predominant signature of DPTs-I: a significant 
change in the dynamical behaviors of physical observables as the control parameter is varied.
Further evidence of the occurrence of DPTs-I as the quenching 
strength passes through the critical value is
provided by a close inspection on the first dip value of $M(t)$, defined as
$M_{\mathrm{dip}}=M(t=\tau_\mathrm{dip})$ with $\tau_\mathrm{dip}$ 
being the time when the first dip present.
The short time behavior of $M(t)$ in Figs.~\ref{TimeMt}(a)-\ref{TimeMt}(c) 
shows that $M_\mathrm{dip}$ has a large
value for $\delta\xi<\delta\xi_c$, while it approaches zero as soon as $\delta\xi>\delta\xi_c$. 
This allows us to take $M_\mathrm{dip}$ as an order parameter of DPT-I 
for the Hamiltonian (\ref{SBECH}).

 \begin{figure}
  \includegraphics[width=\columnwidth]{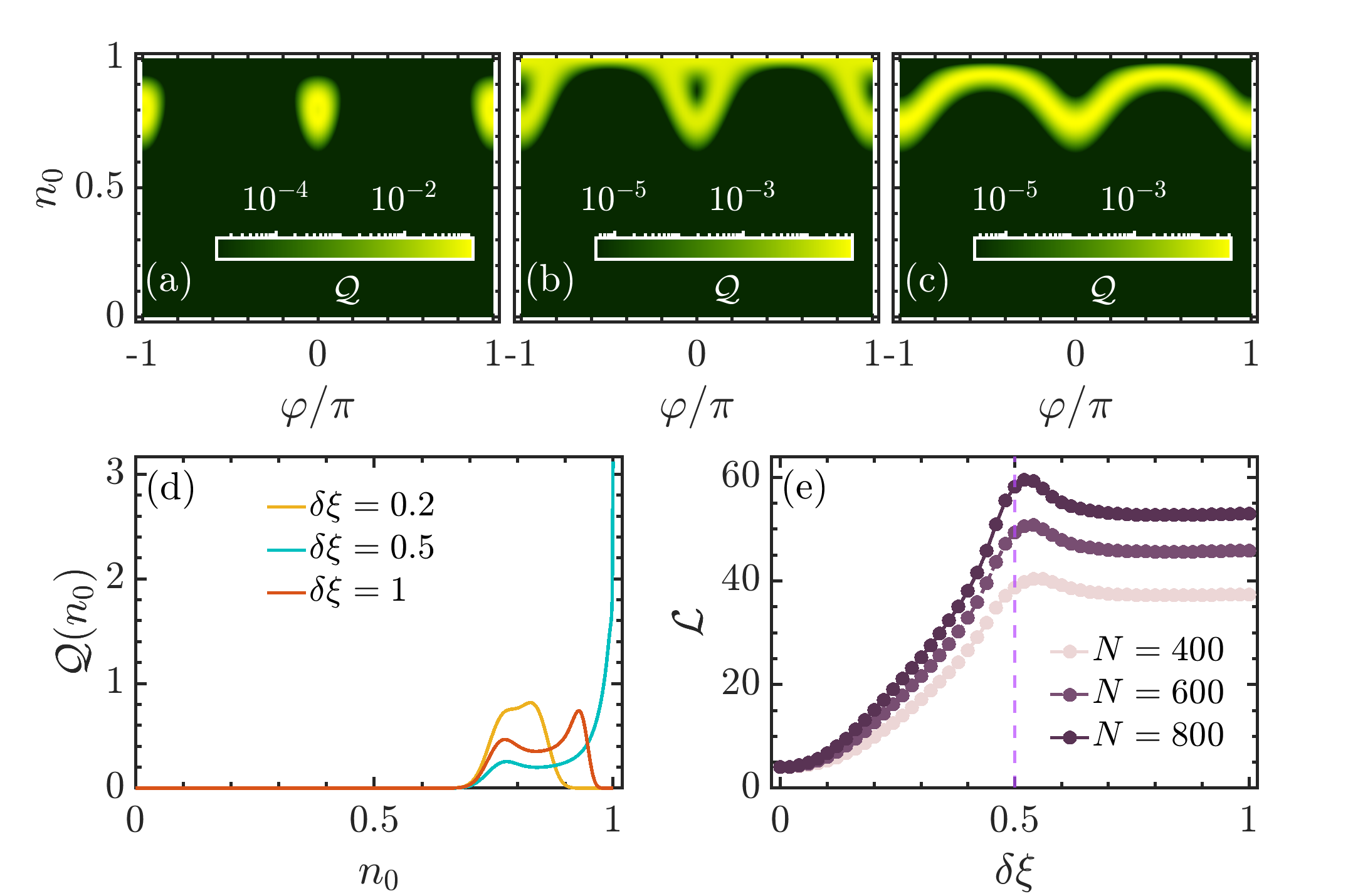}
  \caption{Husimi function $\mathcal{Q}(n_0,\varphi)$ for (a) $\delta\xi=0.2$,
  (b) $\delta\xi=0.5$, and (c) $\delta\xi=1$ with system size $N=600$.
  (d) Marginal distribution of the Husimi function, $\mathcal{Q}(n_0)$ as a function of $n_0$ for several $\delta\xi$ values
  with $N=600$.   
  (e) Phase space localization measure $\mathcal{L}$ as a function of $\delta\xi$ for several system sizes.
  Vertical dashed line marks the critical point, $\delta\xi_c$, of DPT-I.   
  Other parameters: $\xi_i=1$ and $\delta\xi_c=0.5$ obtained from (\ref{CriticalXi}).
  All quantities are dimensionless.}
  \label{HusimiF}
 \end{figure}

Figure \ref{MdipPD}(a) plots $M_\mathrm{dip}$ as a function of $\delta\xi$ for 
several system sizes with $\xi_i=1$.
In the same figure, we also show the corresponded classical result which 
obtains from the first dip in the evolution of $\mathcal{M}_{cl}(t)$.
One see that $M_\mathrm{dip}$ takes a nonzero value and decreases with increasing $\delta\xi$ 
before the critical point has been crossed, whereas it tends to vanish 
as long as the system passes through the critical point. 
The transition in $M_\mathrm{dip}$ is smooth for finite system sizes, rather than 
the abrupt change observed in the classical result.
However, the crossover from $M_\mathrm{dip}\neq0$ to $M_\mathrm{dip}=0$
becomes sharper and tends towards the classical results as the system size increases.
The dependence of $M_\mathrm{dip}$ as a function of $\delta\xi$ and $\xi_i$ is depicted in Fig.~\ref{MdipPD}(b). 
One can clearly see that the phase boundary determined by the dramatic 
change in the behavior of $M_\mathrm{dip}$ is in good agreement with the analytical result. 
We would like to point out that the abrupt change in the behavior of $M_\mathrm{dip}$ can also be recognized as a 
dynamical effect of separatrix in the classical counterpart.  
For $\delta\xi<\delta\xi_c$, the dynamics is restricted within one of two disconnected potential wells with
$\rho_0\neq0$ and $\cos\varphi\neq0$ for all $t$, resulting in positive $M(t)$.  
By contrast, the dynamics can explore two classical wells as long as  $\delta\xi>\delta\xi_c$.
This entails the zero value in $\cos\varphi$, leading to the presence of zero points in the evolution of $M(t)$.

Further signatures of the DPT-I are revealed by performing the scaling analysis.
In Fig.~\ref{MdipPD}(c), we show the scaling of the value of 
$M_\mathrm{dip}$ at $\delta\xi_c$, denoted by $M_\mathrm{dip}^c$, with the system size $N$ for several $\xi_i$ values.
The decay of $M_\mathrm{dip}^c$ with increasing $N$ is well captured by a power law of the form
$M_\mathrm{dip}^c\propto N^{-\eta(\xi_i)}$ with $\eta(\xi_i)\approx0.255$ almost independent of $\xi_i$, 
as demonstrated in the inset in Fig.~\ref{MdipPD}(c).
More scaling properties in $M_\mathrm{dip}$ can be obtained by
considering the finite-$N$ precursor of the critical point $\delta\xi_c(N)$. 
To this end, we identify the location of the minimal value in $\partial M_\mathrm{dip}/\partial(\delta\xi)$ as $\delta\xi_c(N)$. 
The difference between the numerical results and the analytical critical point $\delta\xi_c$ in (\ref{CriticalXi}),
$\Delta_c=|\delta\xi_c(N)-\delta\xi_c|$, evolves as a function of $N$ for several $\xi_i$ values 
is plotted in Fig.~\ref{MdipPD}(d). 
Again, we see the dependence of $\Delta_c$ on the system size is well fitted by the power law decay,
$\Delta_c\propto N^{-\gamma(\xi_i)}$ with
the decay exponent $\gamma(\xi_i)$ increases with increasing $\xi_i$, as seen in the inset of Fig.~\ref{MdipPD}(c).
These results confirm that the system undergoes the DPT-I in the classical limit.

\subsubsection*{Long time average of $M(t)$}

The DPTs-I are also signified by the singular behaviors in the long time average of a certain physical observable. 
Now we focus on the properties of the long time average of $M(t)$, defined as       
\be\label{AvgM}
  \overline{M}=\overline{M(t)}=\lim_{T\to\infty}\frac{1}{T}\int_0^T M(t) dt.
\ee
Substituting $M(t)$ in (\ref{MtEv}) into above equation and note that there is no degeneracies in the spectrum, 
it is easy to find that $\overline{M}$ can be recast as
\be \label{DEavg}
  \overline{M}=\sum_n|c_n|^2M_{nn}=\mathrm{Tr}[\overline{\rho}M],
\ee
where 
\be
   \overline{\rho}=\lim_{T\to\infty}\frac{1}{T}\int_0^T\rho(t)dt=\sum_n|c_n|^2|E_n\ra\la E_n|,
\ee
is the long time averaged density of state.
Notice that this result is in agreement with the diagonal ensemble (DE) \cite{Srednicki1999,Alessio2016}.

 \begin{figure}
  \includegraphics[width=\columnwidth]{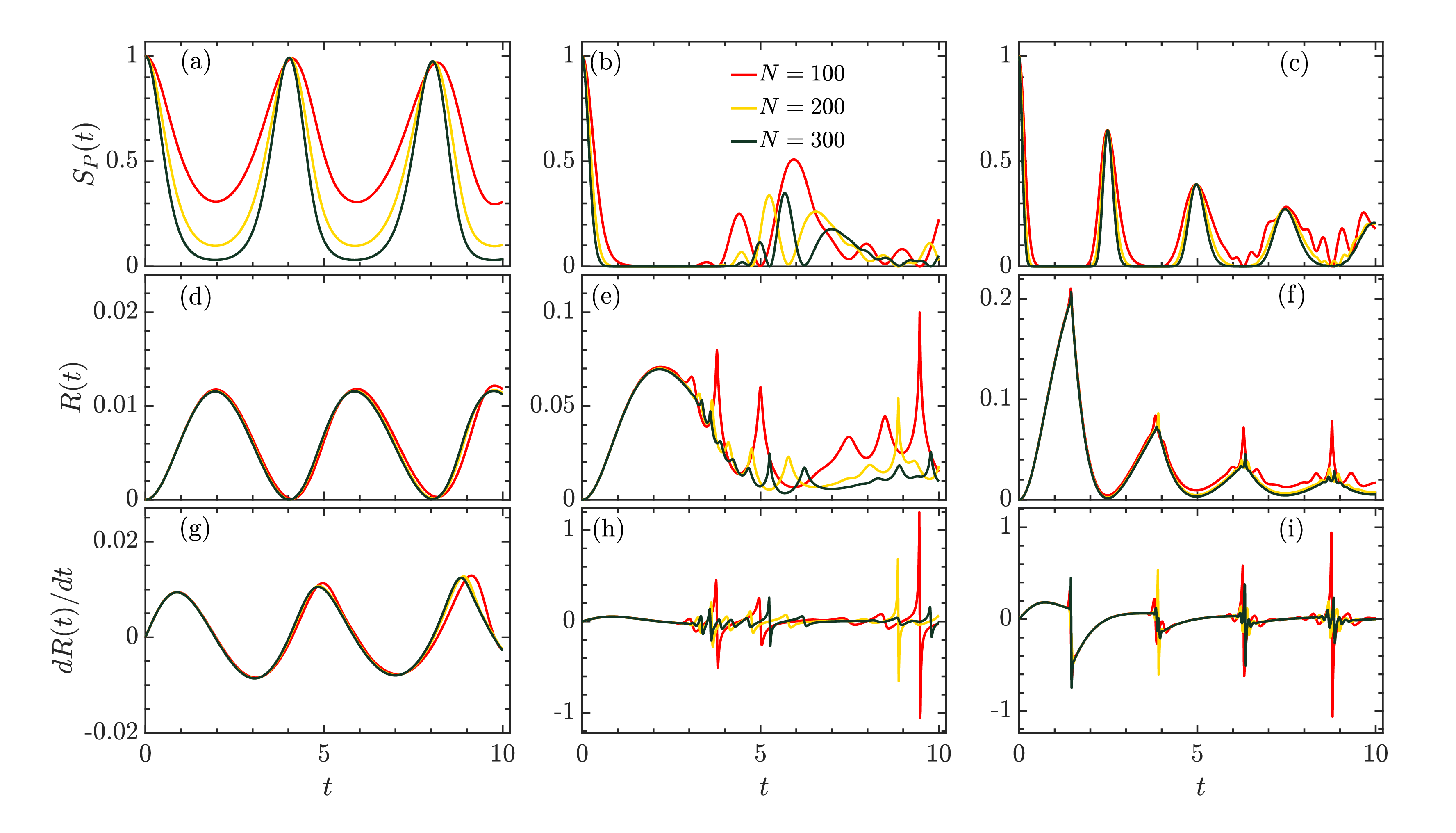}
  \caption{Time evolution of the survival probability $S_P(t)$ for (a) $\delta\xi=0.2$, (b) $\delta\xi=0.5$, and (c) $\delta\xi=1$, 
  and for different system sizes [see the legend in panel (d)].
  (d)-(f): Rate function $R(t)$ corresponding to $S_P(t)$ in panels (a)-(c), respectively.
  (g)-(i): Time derivatives $dR(t)/dt$ for $R(t)$ in panels (d)-(f).
  The legend in panel (b) is used in all panels.   
  Other parameters: $\xi_i=1$ and $\delta\xi_c=0.5$ obtained from (\ref{CriticalXi}).
  The axes in all figures are dimensionless.}
  \label{Ratefs}
 \end{figure}

The calculation of $\overline{M}$ has been conducted for several quenches, letting the evolution of the quenched 
system during $T=1000$.
The final long time averages for a system with $N=1000$ and $\xi_i=1$ are plotted 
as a function of $\delta\xi$ in Fig.~\ref{PDavM}(a). 
Moreover, the long time averages obtained from DE and classical approximation are also
shown in the same figure.  
Irrespective of the method used, we see that the critical point is marked by an abrupt dip in the behavior of $\overline{M}$.
Additionally, an excellent agreement between the numerical and the DE results is clearly visible. 
However, despite the classical result provides a good description of the quantum behavior
in $\delta\xi<\delta\xi_c$ phase, it fails to follow the behavior of $\overline{M}$ for $\delta\xi>\delta\xi_c$.
This is due to the existence of significant quantum correlations in the evolution of $M(t)$, 
as has already been manifested in Fig.~\ref{TimeMt}(c).

The dip exhibited by $\overline{M}$ near the critical point implies that it can be 
taken as the precursor of the critical point of DPT-I.
This can be appreciated in Fig.~\ref{PDavM}(b), where we depcit 
how $\overline{M}$ varies as a function of $\delta\xi$ and $\xi_i$. 
As can be seen, the critical line in (\ref{CriticalXi}) is perfectly reproduced by the dip in $\overline{M}$.
To strengthen above statement, we analyze the scaling of the minimal $\overline{M} $ value, $\overline{M}_{min}$. 
Figure \ref{PDavM}(c) plots $\overline{M}_{min}$ as a function of $N$ for several $\xi_i$ values.
With increasing system size $N$, the decrease of $\overline{M}_{min}$ is well described by the power
law of the form $\overline{M}_{min}\propto N^{-\mu(\xi_i)}$ with $\mu(\xi_i)\approx0.044$. 
An explicit dependence of $\mu(\xi_i)$ on $\xi_i$ is shown in the inset of Fig.~\ref{PDavM}(c).
Hence, the dip in $\overline{M}$ becomes sharp as the system size is increased. 
Moreover, by defining the precursor of the critical point of DPT-I, $\delta\xi_c^m(N)$,
as the position of $\overline{M}_{min}$, we further investigate how the deviation 
between $\delta\xi_c^m(N)$ and $\delta\xi_c$ in (\ref{CriticalXi}), $\Delta_c^m=|\delta\xi_c^m(N)-\delta\xi_c|$, 
evolves as a function of the system size. 
The results for different $\xi_i$ values are demonstrated in Fig.~\ref{PDavM}(d).  
One can see that $\Delta_c^m$ exhibits an obvious power law decay $\Delta_c^m\propto N^{-\zeta(\xi_i)}$,
regardless of the values of $\xi_i$. 
The decay exponent $\zeta(\xi_i)$ is approximately independent of $\xi_i$ 
and given by $\zeta(\xi_i)\approx0.7805$, as visualized in the inset of Fig.~\ref{PDavM}(d).
This confirms that the emergence of DPT-I is characterized by the dip in the long time average of $M(t)$.

 \begin{figure}
  \includegraphics[width=\columnwidth]{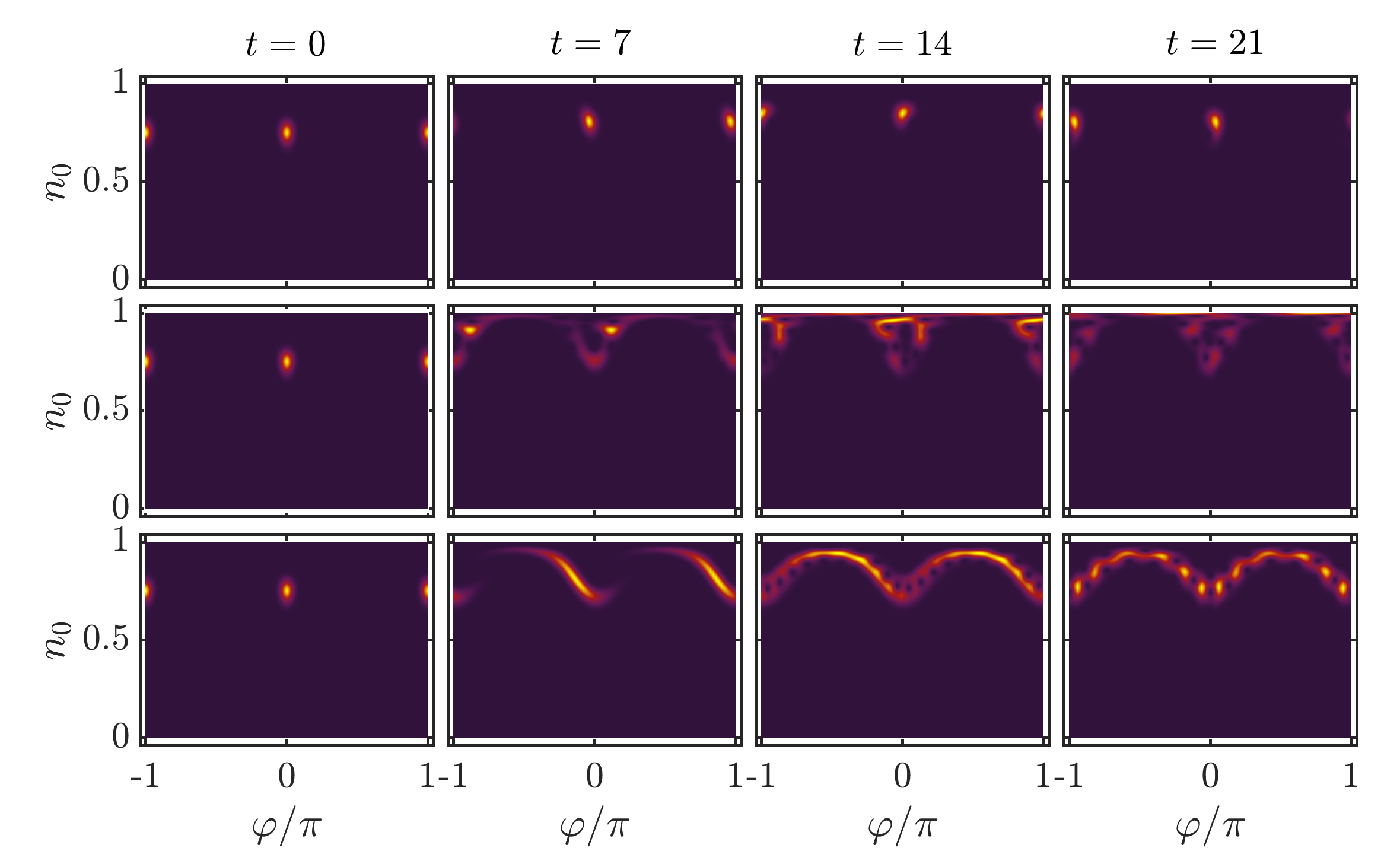}
  \caption{Snapshots of the evolved Husimi function $\mathcal{Q}_t(n_0,\varphi)$ at different time steps 
  with $\delta\xi=0.2, 0.5$, and $1$ (from top to bottom rows).  
  Other parameters: $N=300$ and $\xi_i=1$ which yields $\delta\xi_c=0.5$ [cf.~Eq.~(\ref{CriticalXi})].
  The axes in all figures are dimensionless.}
  \label{Hsftime}
 \end{figure}

The abrupt decrease exhibited by $\overline{M}$ around the critical point is also 
a manifestation of the separatrix in the classical dynamics. 
Classically, since the separatrix associated with the saddle point, 
the initial state evolves into a rather delocalized state along the separatrix.
Consequently, both classical and quantum evolutions of the order parameter are able to
 reach very small values, giving rise to a remarkable dip in 
 $\overline{M}$ near the critical point of DPT-I.    
To illustrate how the extension of the long time averaged state depends 
on the quenching strength, we consider the Husimi function \cite{Husimi1940}.
The Husimi function is the Gaussian smooth of the well known Wigner function \cite{Wigner1932} 
and provides a useful tool for studying quantum-classical correspondence in various systems.
In the classical limit, the Husimi function behaves as a classical probability distribution in phase space 
and evolves according to the Liouville equation \cite{Altland2012a,Altland2012b}.
For the state $\overline{\rho}$, the Husimi function can be written as
\be
   {\mathcal{Q}}(n_0,\varphi)=\la\bm{\alpha}|\overline{\rho}|\bm{\alpha}\ra=\sum_n|c_n|^2|\la\bm{\alpha}|E_n\ra|^2,
\ee
where $|\bm{\alpha}\ra$ is the coherent states in (\ref{Coherents}) with $N_{-1}=N_1$. 
The normalization condition of $\mathcal{Q}(n_0,\varphi)$ is given by
\be
   \frac{N+1}{2\pi}\int dn_0d\varphi\mathcal{Q}(n_0,\varphi)=1.
\ee

Density plots of $\mathcal{Q}(n_0,\varphi)$ for several $\delta\xi$ values in a 
system with $N=1000$ and $\xi_i=1$ are shown in Figs.~\ref{HusimiF}(a)-(c).
The largest extension of $\overline{\rho}$ in the phase space for the 
critical quench $\delta\xi_c=0.5$ is clearly visible. 
In particular, we see that the Husimi function for $\delta\xi_c=0.5$ case 
shows an obvious distribution the line of $n_0=1$. 
This is more evidenced in the behavior of the marginal distribution of the Husimi function, defined as
\be
   \mathcal{Q}(n_0)=\sqrt{\frac{N+1}{2\pi}}\int d\varphi\mathcal{Q}(n_0,\varphi),
\ee 
with normalization condition
\be
  \sqrt{\frac{N+1}{2\pi}}\int dn_0\mathcal{Q}(n_0)=1.
\ee
It is the projection of the Husimi function in the $n_0$-space. 
Figure \ref{HusimiF}(d) shows how $\mathcal{Q}(n_0)$ varies as a function of $n_0$ for 
the same values of $\delta\xi$ as in Figs.~\ref{HusimiF}(a)-(c).
The condensation of the Husimi function along $n_0=1$ observed in Fig.~\ref{HusimiF}(b)
is unambiguously confirmed by the highest peak 
in $\mathcal{Q}(n_0)$ at $n_0=1$ for $\delta\xi=\delta\xi_c=0.5$ case.

The degree of extension of the state $\overline{\rho}$ can be quantified 
by the phase space localization measure, defined as
\be
  \mathcal{L}=\left[\frac{N+1}{2\pi}\int dn_0d\varphi\mathcal{Q}^2(n_0,\varphi)\right]^{-1}.
\ee
It is the inverse of the second momentum of the Husimi function 
and can be considered as the participation ratio of $\overline{\rho}$ in the phase space. 
The definition of $\mathcal{L}$ implies that it varies in the range $\mathcal{L}\in[0,N+1]$.
For the extremely localized state, it is identical to a single point in phase space.
This means $\mathcal{L}\simeq1/(N+1)$ which vanishes as $N\to\infty$. 
However, for the state that uniformly covers the phase space, we have $\mathcal{Q}(n_0,\varphi)=1/(N+1)$,
resulting in $\mathcal{L}=N+1$ in this case.
Hence, the growth of the value of $\mathcal{L}$ implies the increase of the degree of 
delocalization of $\overline{\rho}$ in phase space.
In Fig.~\ref{HusimiF}(e), we plot $\mathcal{L}$ as a function of $\delta\xi$ 
for several system sizes with $\xi_i=1$ which yields $\delta\xi_c=0.5$ [cf.~Eq.~(\ref{CriticalXi})].
Irrespective of the system size, the localization measure exhibits a peak in the vicinity of the critical point,
indicating that the state $\overline{\rho}$ has the highest degree of delocalization near the critical point,
as seen in Fig.~\ref{HusimiF}(b).
This further verifies that the DPT-I is a dynamical consequence of the separatrix in the underlying classical systems.

 \begin{figure}
  \includegraphics[width=\columnwidth]{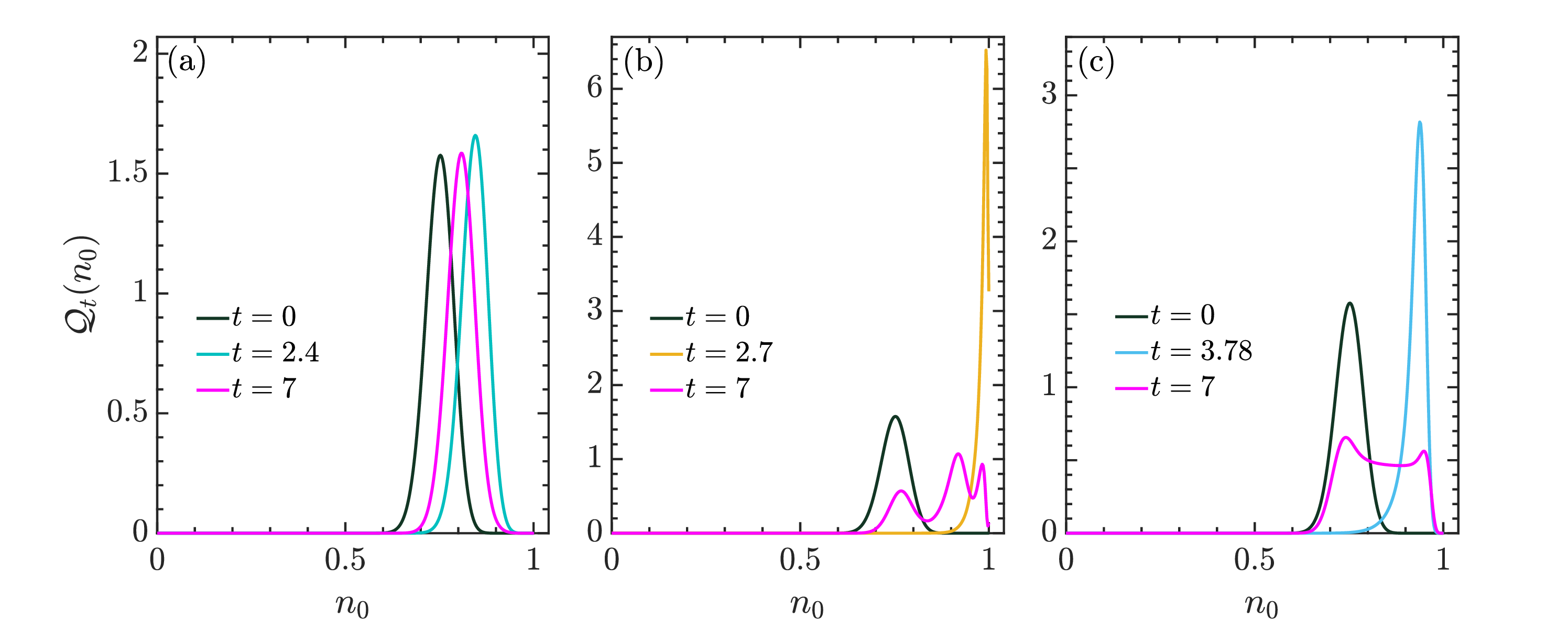}
  \caption{Marginal distribution, $\mathcal{Q}_t(n_0)$, of the time evolved Husimi function at different times
  for (a) $\delta\xi=0.2$, (b) $\delta\xi=0.5$, and (c) $\delta\xi=1$.  
  Other parameters: $N=300$ and $\xi_i=1$, so that $\delta\xi_c=0.5$ [cf.~Eq.~(\ref{CriticalXi})].
  The axes in all figures are dimensionless.}
  \label{Mhft}
 \end{figure}

\subsection{DPTs-II: cusps in survival probability} \label{thirdB}

The second kind of DPTs, DPTs-II, are usually triggered by different physical origin than DPTs-I. 
Even though the connections between this two kinds of DPTs have been revealed in several 
systems from different aspects
\cite{Lerose2019,Zunkovic2016,Zunkovic2018,LangJ2018,Sehrawat2021,Weidinger2017,JFrank2018,Hashizume2022}, 
such as the the links between the zeros in the order parameters of DPTs-I
and the critical points of DPTs-II, a general understanding on the mechanism of DPTs-II remains unknown.
   
In the pioneering work \cite{Heyl2018}, a DPT-II is defined as the onset of singularities 
in the evolution of the initial state survival probability rate function at critical times.  
For a given initial state $|\psi_0\ra$, its survival probability and associated rate function are defined as
\be\label{Qusrt}
   S_P(t)=|\la\psi_0|e^{-iH_ft}|\psi_0\ra|^2,\ R(t)=-\frac{1}{N}\ln S_P(t),
\ee
where $N$ denotes the system size.
The definition of DPTs-II is based on the resemblance of the amplitude of $S_p(t)$ 
to the thermal partition function $\mathcal{Z}(\beta)=\mathrm{Tr}e^{-\beta H}$.
This can be seen by considering the amplitude of $S_P(t)$ as a complex function with variable $z=it$, so that
$\mathcal{A}_P(z)=\la\psi_0|e^{-zH}|\psi_0\ra=\mathrm{Tr}e^{-zH}$. 
Thus, one can extend the concept of equilibrium phase transition, which signifies as singularities in the free 
energy density $F=-(1/\beta N)\ln\mathcal{Z(\beta)}$, to DPTs-II by 
identifying the rate function $R(t)$ as the dynamical free energy density. 
Analogy to the thermal phase transition, the occurrence of a DPT-II is 
captured by nonanalytical in $R(t)$ at certain times.
Moreover, the correspondence of $\mathcal{A}_P(z)$ and $\mathcal{Z}(\beta)$ also
motives numerous investigations of the relationship between 
nonequilibrium and equilibrium phase transitions \cite{Heyl2018,Karrasch2013,Vajna2014,Schmitt2015,Zvyagin2016}.

Previous works have been demonstrated that the presence of a DPT-II requires quench the control parameter
across the ground state quantum phase transition \cite{Heyl2013,Heyl2018,Heyl2014}. 
However, the recent studies in the long-range interacting and collective systems 
were found that DPTs-II can be divided into two distinct scenarios \cite{Corps2022,Zauner2017,Homrighausen2017}.  
The first one, on the one hand, is dubbed as regular dynamic phase, characterized by 
appearing of the first nonanalytical cusp always before the 
first minimal value of the rate function \cite{Homrighausen2017}.
The critical quench that leads to the regular 
dynamical phase are independent of the ground state phase transition \cite{Sciolla2011}. 
On the other hand, a DPT-II can also emerge even for the quenches that are not 
crossed the ground state quantum phase transition, 
resulting in the so called anomalous dynamical phase \cite{Zauner2017,Homrighausen2017}. 
In contrast to the regular dynamical phase, a main signature of the anomalous dynamical phase 
is the cusps occurring only after the first minimum of the rate function \cite{Corps2022,Homrighausen2017}. 
The following of this subsection devotes to explore the DPT-II in spinor BEC from 
both quantum and semiclassical perspectives.

 \begin{figure}
  \includegraphics[width=\columnwidth]{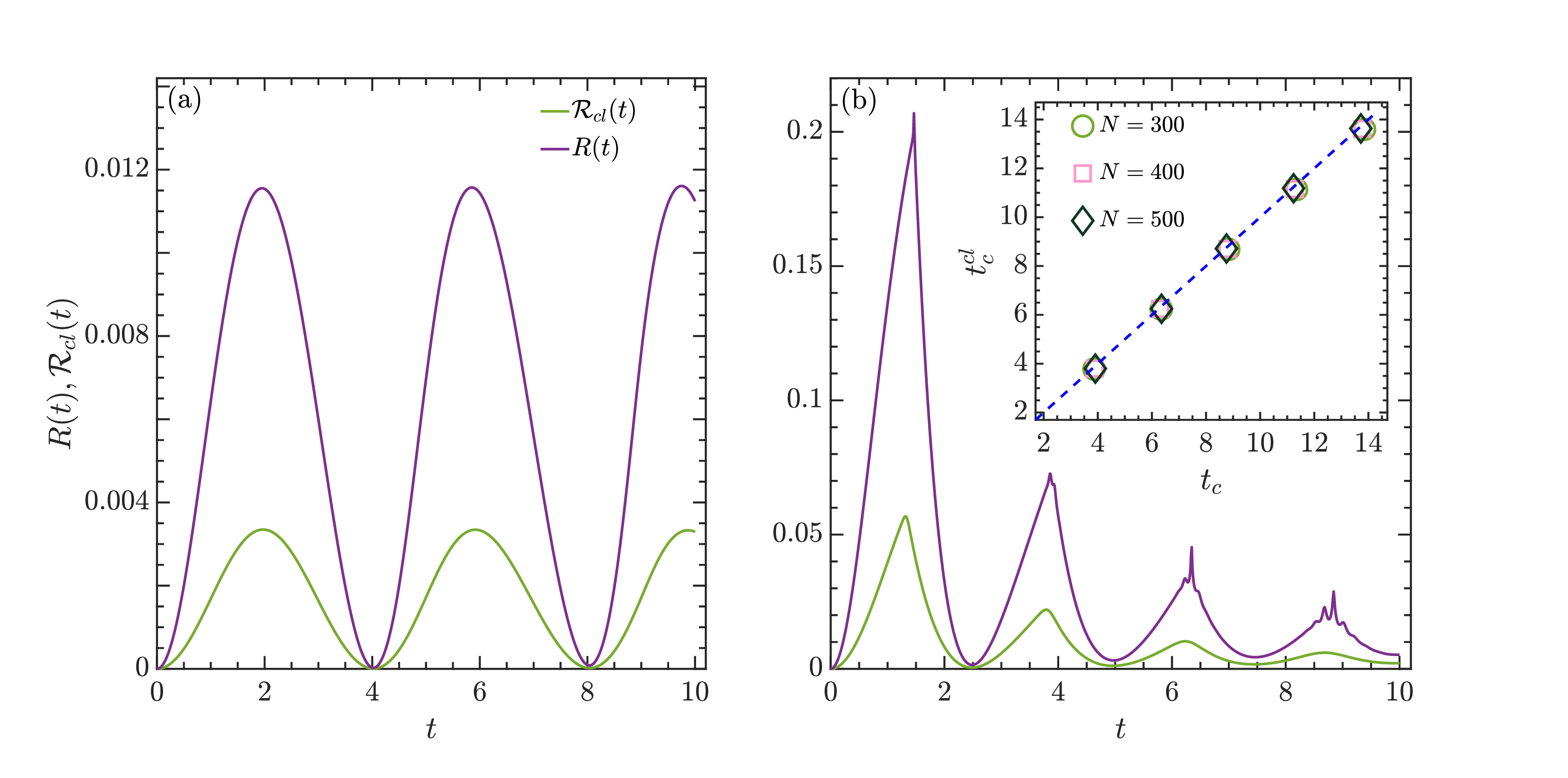}
  \caption{Semiclassical rate function $\mathcal{R}_{cl}(t)$ in (\ref{Scsrt}), 
  along with quantum counterpart $R(t)$ in (\ref{Qusrt})
  for (a) $\delta\xi=0.2$, (b) $\delta\xi=1$. 
  Inset in panel (b) plots the critical times $t_c^{cl}$ corresponding to kinks in $\mathcal{R}_{cl}(t)$ 
  and $t_c$ obtained from the nonanalytical behavior in $R(t)$ for several system sizes. 
  The blue dashed line denotes the linear fitting with the form $t_c^{cl}=t_c$.  
  Other parameters: $N=300$ and $\xi_i=1$, so that $\delta\xi_c=0.5$ [cf.~Eq.~(\ref{CriticalXi})].
  All quantities are dimensionless.}
  \label{ClassicalRt}
 \end{figure}

In Figs.~\ref{Ratefs}(a)-\ref{Ratefs}(c), we plot the evolution of $S_P(t)$ for 
different $\delta\xi$ values and several system sizes with $\xi_i=1$, which results in $\delta\xi_c=0.5$.
If $\delta\xi<\delta\xi_c$, as seen in Fig.~\ref{Ratefs}(a), $S_P(t)$ behaves as a simply smooth 
and periodic function of time and it always takes positive values during the evolution.
However, $S_P(t)$ exhibits a completely different behavior when $\delta\xi>\delta\xi_c$, 
as exemplified in Fig.~\ref{Ratefs}(c).
Although it still shows a regular oscillation, the magnitude of $S_P(t)$ bears a time decay in this case.
In particular, we see that $S_P(t)$ vanishes periodically with time.
The vanish of $S_P(t)$ is also presented for the critical quench, 
but no periodic behavior can be observed in the evolution of $S_P(t)$, as illustrated in Fig.~\ref{Ratefs}(b).

The observed features of $S_P(t)$ indicate the existence of DPT-II for the case of $\delta\xi\geq\delta\xi_c$. 
This is clearly visible in Figs.~\ref{Ratefs}(d)-\ref{Ratefs}(f), where we display the corresponded rate 
functions of $S_P(t)$ in Figs.~\ref{Ratefs}(a)-\ref{Ratefs}(c). 
The obvious kinks in $R(t)$ at certain critical times $t_c$ in 
Figs.~\ref{Ratefs}(e) and \ref{Ratefs}(f) are strikingly contrasted to the smooth behavior shown in Fig.~\ref{Ratefs}(d) 
and unveil the occurrence of DPT-II in the system when $\delta\xi\gtrsim\delta\xi_c$. 
Moreover, due to the first nonanalytical point appears before the the first minimum of $R(t)$, 
the phase of DPT-II is regular.
The presence of DPT-II for $\delta\xi\gtrsim\delta\xi_c$ is further revealed 
as the nonsmoothness in the time derivative of $R(t)$, as demonstrated in 
Figs.~\ref{Ratefs}(g)-\ref{Ratefs}(i), where we plot $dR(t)/dt$ as a function of time.

More insights on the properties of DPT-II can be gained by considering 
the evolution of the quantum state in the classical phase space. 
To this end, let us focus on the Husimi function of the evolved state, calculated as
\be
  \mathcal{Q}_t(n_0,\varphi)=\la\bm{\alpha}|\rho_t|\bm{\alpha}\ra
    =\left|\sum_ke^{-iE_k^ft}\la\bm{\alpha}|E_k^f\ra\la E_k^f|\bm{\alpha}\ra\right|^2,
\ee
where $\rho_t$ is the density of the evolved state given in Eq.~(\ref{densityt}).
The evolution of $\mathcal{Q}_t(n_0,\varphi)$ at different times for several values 
of $\delta\xi$ has been plotted in Fig.~\ref{Hsftime} for a system with $N=300$ and $\xi_i=1$.
In this case, we have $\delta\xi_c=0.5$, according to Eq.~(\ref{CriticalXi}).
We clearly see that the specific evolution process of $\mathcal{Q}_t(n_0,\varphi)$ is 
determined by the strength of the quench.
For small $\delta\xi$, since the quenched system without 
enough energy to overcome the saddle point,
the evolution of $\mathcal{Q}_t(n_0,\varphi)$ is confined within 
two disjointed phase space regions 
and exhibits a periodically rotation around the initial state, as illustrated in the top row of Fig.~\ref{Hsftime}.  
This leads to the regular oscillation behavior in survival probability with nonzero values.
Hence, the rate $R(t)$ is a simply periodic and smooth function.
This explains the absence of DPT-II when $\delta\xi<\delta\xi_c$. 
In contrast, as soon as $\delta\xi\gtrsim\delta\xi_c$, the quenched system acquires enough energy 
so that it enables to explore both regions freely, as observed in the middle and bottom rows of Fig.~\ref{Hsftime}.
This means the quenched state can exhibit a zero overlap with initial state at certain times, 
leading to zero values in the survival probability. 
As a consequence, the rate $R(t)$ behaves as a nonanalytical function 
with kinks at different critical times $t_c$ for $\delta\xi\gtrsim\delta\xi_c$, 
and thus the onset of DPT-II. 
These results are more clearly visible in Fig.~\ref{Mhft}, where we plot the 
marginal distribution of $\mathcal{Q}_t(n_0,\varphi)$ at several time steps for different $\delta\xi$ values.
The overlap between $\mathcal{Q}_{t\neq0}(n_0)$ and $\mathcal{Q}_{t=0}(n_0)$ 
is alway finite for $\delta\xi<\delta\xi_c$, 
while it allows to take zero or very tiny value as long as $\delta\xi\gtrsim\delta\xi_c$.

 \begin{figure}
  \includegraphics[width=\columnwidth]{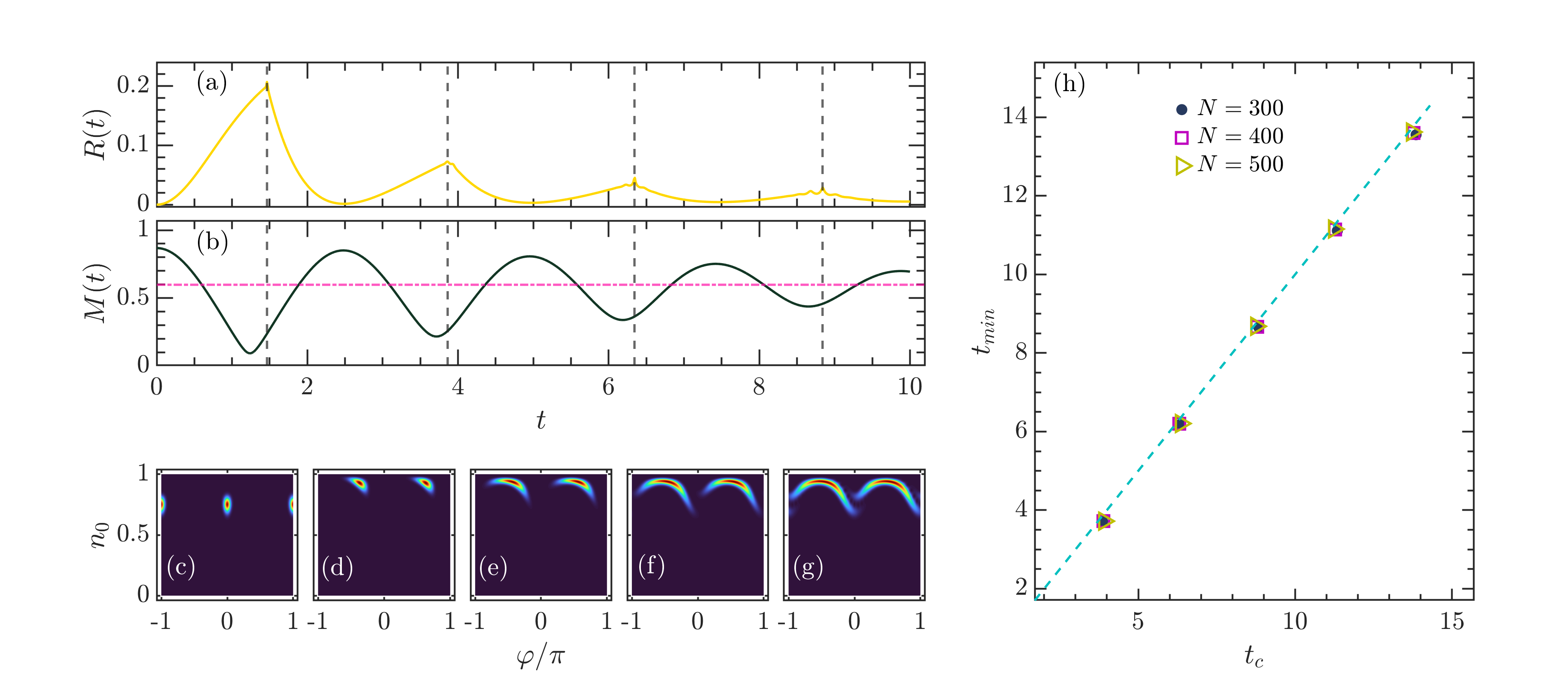}
  \caption{(a) Rate function $R(t)$ and (b) the evolution of $M(t)$ for $\delta\xi=1$ with $N=300$ 
  and $\xi_i=1$. In this case, one can find the critical quench is given by $\delta\xi_c=0.5$ [cf.~Eq.~(\ref{CriticalXi})].
  The vertical dashed lines in panels (a) and (b) signal the critical times of DPT-II,
  while the horizontal dot-dashed line in (b) marks the long time averaged value of $M(t)$, 
  obtained from Eq.~(\ref{DEavg}). 
  (c)-(g): Snapshots of the Husimi function at (c) $t=0$, (d) $t=t_{c,1}=1.4632$, (e) $t=t_{c,2}=3.860$,
  (f) $t=t_{c,3}=6.345$, and (g) $t=t_{c,4}=8.840$ for the same parameters as in panels (a) and (b).   
  (h) Times $t_{min}$, associated with the local minimal of $M(t)$ and the critical times $t_c$ of DPT-II
  for several system sizes with $\xi_i=1$ and $\delta\xi=1$.
  The dashed line represents the linear function of the form $t_{mim}=t_c$.
  The axes in all figures are dimensionless.}
  \label{RelationRM}
 \end{figure}

To make aforementioned points strong, we study the semiclassical counterpart 
of the survival probability and associated rate function.
By considering the normalization condition of the Husimi function, we define them as
\be \label{Scsrt}
  \mathcal{S}_{cl}(t)=\frac{N+1}{2\pi}\int dn_0d\varphi\left[\mathcal{Q}_t(n_0,\varphi)
   \mathcal{Q}_0(n_0,\varphi)\right]^{1/2},\quad
 \mathcal{R}_{cl}(t)=-\frac{1}{N}\ln\mathcal{S}_{cl}(t).
\ee
Here, the definition of $\mathcal{S}_{cl}(t)$ guarantees 
$\mathcal{S}_{cl}(t)\leq1, \forall t$, such as the quantum survival probability.
The features observed in the evolution of $\mathcal{Q}_t(n_0,\varphi)$ and associated marginal distribution in
Figs.~\ref{Hsftime} and \ref{Mhft} imply that $\mathcal{R}_{sc}(t)$ 
should be smooth function when $\delta\xi<\delta\xi_c$, whereas it would 
become nonanalytical with kinks at certain times for $\delta\xi>\delta\xi_c$ case.
This is verified in Fig.~\ref{ClassicalRt}, where $\mathcal{R}_{sc}(t)$ along with its quantum counterpart $R(t)$ 
for different $\delta\xi$ values are plotted. 
Irrespective of $\delta\xi$ value, the resemblance in the behaviors 
of $\mathcal{R}_{cl}(t)$ and $R(t)$ can be clearly seen.
To quantitatively confirm the equivalence between $\mathcal{R}_{cl}$ and $R(t)$, 
we compare the critical times $t_c^{cl}$, associated with the kinks in $\mathcal{R}_{cl}(t)$ 
to the times $t_c$ that are extracted from the nonanalytical points in the evolution of $R(t)$. 
In the inset of Fig.~\ref{ClassicalRt}(b), we plot $t_c^{cl}$ as a function of $t_c$ for several system sizes.  
The dependence of $t_c^{cl}$ on $t_c$ is well captured by a linear function of the form $t_c^{cl}=t_c$,
demonstrating the usefulness of the semiclassical approach in studying DPT-II. 
Moreover, the coincidence between $t_c^{cl}$ and $t_c$ further corroborates that the presence of DPT-II
can also be attributed to the existence of the separatix in the corresponding classical dynamics.

As the separatrix in the classical system links to both DPT-I and DPT-II, 
it is natural to ask what is the relationship between them. 
Although the pervious studies in the long-range interacting spin models have shown that   
the presence of DPT-II is associated with the vanish of the order parameter of DPT-I 
\cite{Zauner2017,Zunkovic2018,JFrank2018,Weidinger2017,Homrighausen2017},
a general connection between two kinds of DPTs remains an open question.
Here, at the end of this section, let us address this question in 
spin-$1$ BEC system by investigating whether the nonanalytical times in $R(t)$ are linked to
the times when the evolution of the order parameter $M(t)$ of DPT-I takes local minima. 
In Fig.~\ref{RelationRM}(a), we plot $R(t)$ for $\xi_i=1$ and $\delta\xi=1$ with $N=300$,
and Fig.~\ref{RelationRM}(b) shows the evolution of $M(t)$ for same parameters.
One can see that the nonanalytical points in $R(t)$, i.~e., critical times $t_c$, 
are close to the times $t_{min}$, at which $M(t)$ present local minima. 
One can expected the the observed deviations between $t_c$ and $t_{min}$ can be 
suppressed by increasing system size.
Moreover, the Husimi functions for initial time and each critical time $t_c$ 
are also depicted in Figs.~\ref{RelationRM}(c)-\ref{RelationRM}(f).
As we can see, the overlap of the evolved Husimi function with the initial one 
is zero or very tiny at the critical times $t_c$ of DPT-II.  
Meanwhile, the Husimi function at each critical time $t_c$ is 
highly condensed around $n_0=1$, resulting in the local minima in $M(t)$. 
These features of the Husimi function explains why the critical times of DPT-II are 
correlated with the the local minima of $M(t)$.
To make such connection more clearer, we compare the critical times $t_c$ to the 
local minima times $t_{min}$ of $M(t)$ in Fig.~\ref{RelationRM}(h).
We see that $t_{min}$ exhibits an obvious linear dependence on $t_c$ and 
it converges to the critical times with increasing system size. 
Hence, the local minima in $M(t)$ indeed reveals the presence of DPT-II.

\section{Conclusions} \label{fourth} 

In this work, a detailed investigation of the DPTs in
a many body quantum system, namely the celebrated spin-$1$ BEC, has been performed.
As the spin-$1$ BEC can be controlled in a highly precision and has rich phases, 
it can be used as a suitable platform to investigate the DPTs.
We have shown that two types of DPTs are presented as the control parameter 
has been quenched through the critical value.
A semiclassical analysis results in an analytical expression of the critical quench 
and reveals that both types of DPTs can be recognized as the dynamical consequence 
of the separatrix in the underlying classical dynamics.
   
The characterizations of two types of DPTs have been scrutinized 
by various quantum and semiclassical properties. 
Specifically, we uncover the signatures of DPT-I by analyzing the 
time evolution of the order parameter, $M(t)$, and its long time average features 
for both quantum and semiclassical systems.
Semiclassically, depending on the quenching strength, the semiclassical dynamics is either 
locked within two disconnected regions in the phase space, or explore the whole phase space.
This leads to a dramatic change in the quantum dynamics of the order parameter.
We have shown that the quantum dynamics follows the semiclassical counterpart 
up to a certain time for the quenches that below the critical value. 
However, the agreement between quantum and semiclassical dynamics disappears quickly 
when the quench strength larger than the critical value. 
Both quantum and semiclassical dynamics exhibit particular behaviors at the critical quench.
The presence of DPT-I is more clearly revealed by the first dip of $M(t)$.
We have shown that the first dip of $M(t)$ behaves as an order parameter of DPT-I. 
It has nonzero value if the quench strength below the critical value, 
while it becomes zero in the semiclassical limit once the quench strength above the critical value. 
By performing a scaling analysis, we have confirmed that the first dip of $M(t)$ 
is indeed an appropriate order parameter of DPT-I for our system.
We have also examined the properties of the long time averaged order parameter, $\overline{M}$, 
and demonstrated that the dependence of $\overline{M}$ on the quenching strength 
is well captured by the diagonal ensemble. 
In contrast to the case of the first dip in $M(t)$, 
DPT-I appears as a  sudden dip in the behavior of $\overline{M}$.
By employing the Husimi function, we have shown that the presence of the dip in 
$\overline{M}$ around the critical point is due to the extension 
of the evolved state over the separatrix.

Regarding to DPT-II, we have shown that it can only happen 
when the quenching strength above the critical value, while it disappears 
as long as the quenching strength smaller than the critical value. 
The occurrence of DPT-II is signified by the kinks 
in the rate function of the initial state survival probability at different critical times.
A semiclassical explanation of the emergence of DPT-II has been 
discussed by means of Husimi function. 
We have found that the onset of DPT-II depends on whether the 
quenched system has enough energy to straddle the saddle point freely.
For quenching strengths that below the critical value, the quenched system 
does not have sufficient energy to surpass the saddle point. 
As a result, the evolution of the system is confined within the initial position, 
leading to a nonzero overlap between the evolved and initial states
and the absence of DPT-II.
On the contrary, as soon as the quench strength above the critical value,
the quenched system acquires enough energy so that it can pass through the saddle point. 
This means that the evolved state will exhibit a zero overlap with the initial state at certain times.
Thus, the rate function undergoes a nonanalytical behavior at different times, 
 indicating the presence of DPT-II.
 The correctness of this semiclassical picture of DPT-II 
 is further verified by the agreement between the behaviors 
 of the quantum and semiclassical rate functions.
 We can therefore conclude that both DPTs-I and DPTs-II in the spin-$1$ BEC are 
 triggered by the separatrix in the corresponding classical system. 
 This conclusion is further evidenced by the direct link between the critical times of DPT-II 
 and the times at which $M(t)$ presents local minima.
 A numerical calculation has been suggested that these 
 two times are consistent with each other in the semiclassical limit.

 A continuation of this work is to analyze the scaling properties of DPTs-II.
 The classification of DPTs-II remains an open question, despite it is 
 valuable for understanding DPTs-II \cite{Heyl2015,Bandy2021,Bhat2023}. 
 It would also be interesting to extend our present analysis for ferromagnetic condensate 
 to the antiferromagnetic case, which undergoes a first order ground state 
 quantum phase transition as the control parameter passes through the critical value. 
 The DPTs-I in the antiferromagnetic spinor condensate has been studied
 in recent works \cite{YangHX2019,HuangY2022}, but the studies of DPTs-II is still lacking. 
 Moreover, it is worth stressing that our work verifies the usefulness 
 of the semiclassical approach in comprehending DPTs.
 Hence, we hope that the present work can motivate other semiclassical studies of DPTs 
 in many body quantum systems that have the well-defined semiclassical limit.
 As a final remark, owing to the spin-$1$ BEC is a highly controllable platform and
 the emergence of various advanced techniques that enable to measure the density matrices,
 we expect that our findings could stimulate more experimental researches on the properties of DPTs.

 \acknowledgments

 This work was supported by the National Science Foundation of China under Grant No.~11805165;
 the Zhejiang Provincial Nature Science Foundation under Grant Nos.~LQ22A040006 and LY20A050001.   
Q.~W. acknowledges support from the Slovenia Research and Innovation Agency (ARIS) under the
Grant Nos.~J1-4387 and P1-0306.

 \begin{figure}
  \includegraphics[width=\columnwidth]{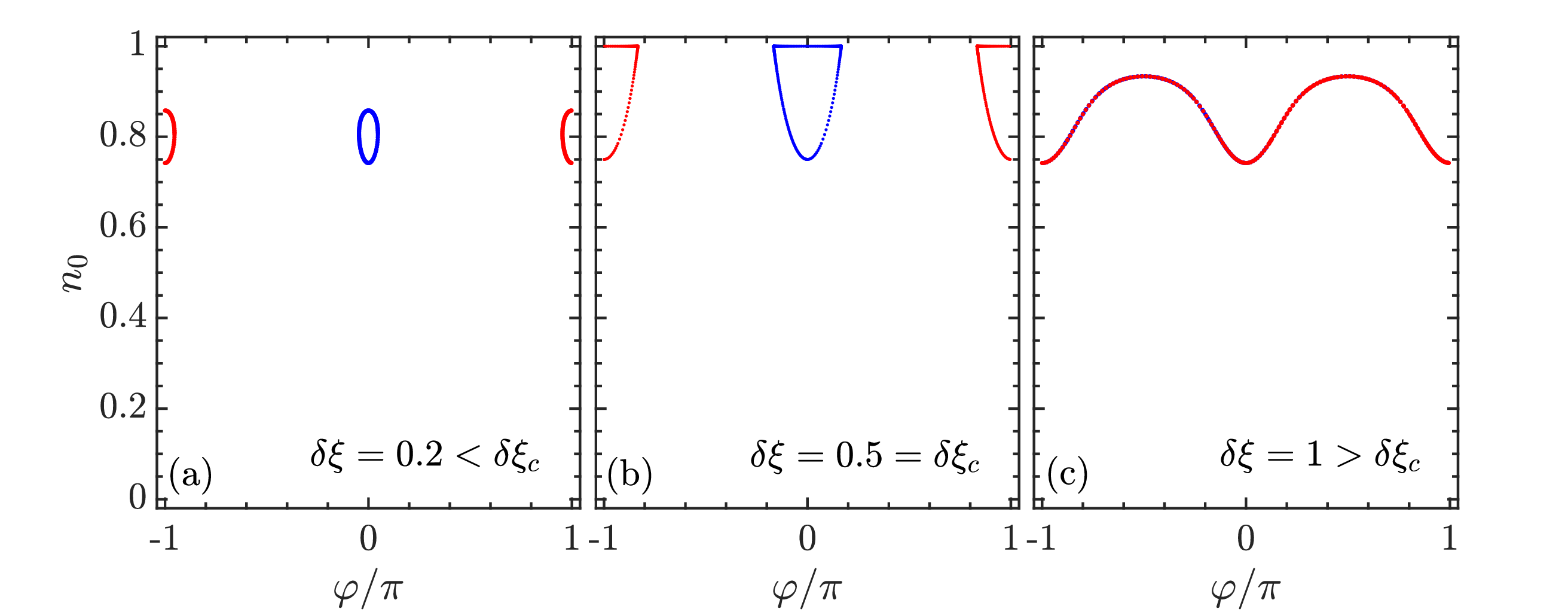}
  \caption{Snapshots of the semiclassical dynamics of the spin-$1$ BEC 
  for several $\delta\xi$ values, obtained from the classical 
  equations of motion (\ref{EOM}) in the main text. 
  Here, the system is evolved up to $t=22$ with initial condition
  $(\cos\varphi_i,n_{0,i})=[\pm1,(2+\xi_i)/4]$.
  The dynamics of the initial condition 
  $\cos\varphi_i=1$ is plotted by blue dots, while
  red dots correspond to the case of $\cos\varphi_i=-1$.  
  Other parameter: $\xi_i=1$, 
  so that $\delta\xi_c=0.5$ [cf.~Eq.~(\ref{CriticalXi})]. 
  The axes in all figures are dimensionless.}
  \label{Cdyt}
 \end{figure}

\appendix

\section{Semiclassical dynamics of the spin-$1$ BEC} \label{Appa}

To get a better understanding of the semiclassical origins of DPTs and 
provide further evidence of quantum-classical correspondence of nonequilibrium dynamics,
we consider the semiclassical dynamics in this appendix. 

 \begin{figure}
  \includegraphics[width=\columnwidth]{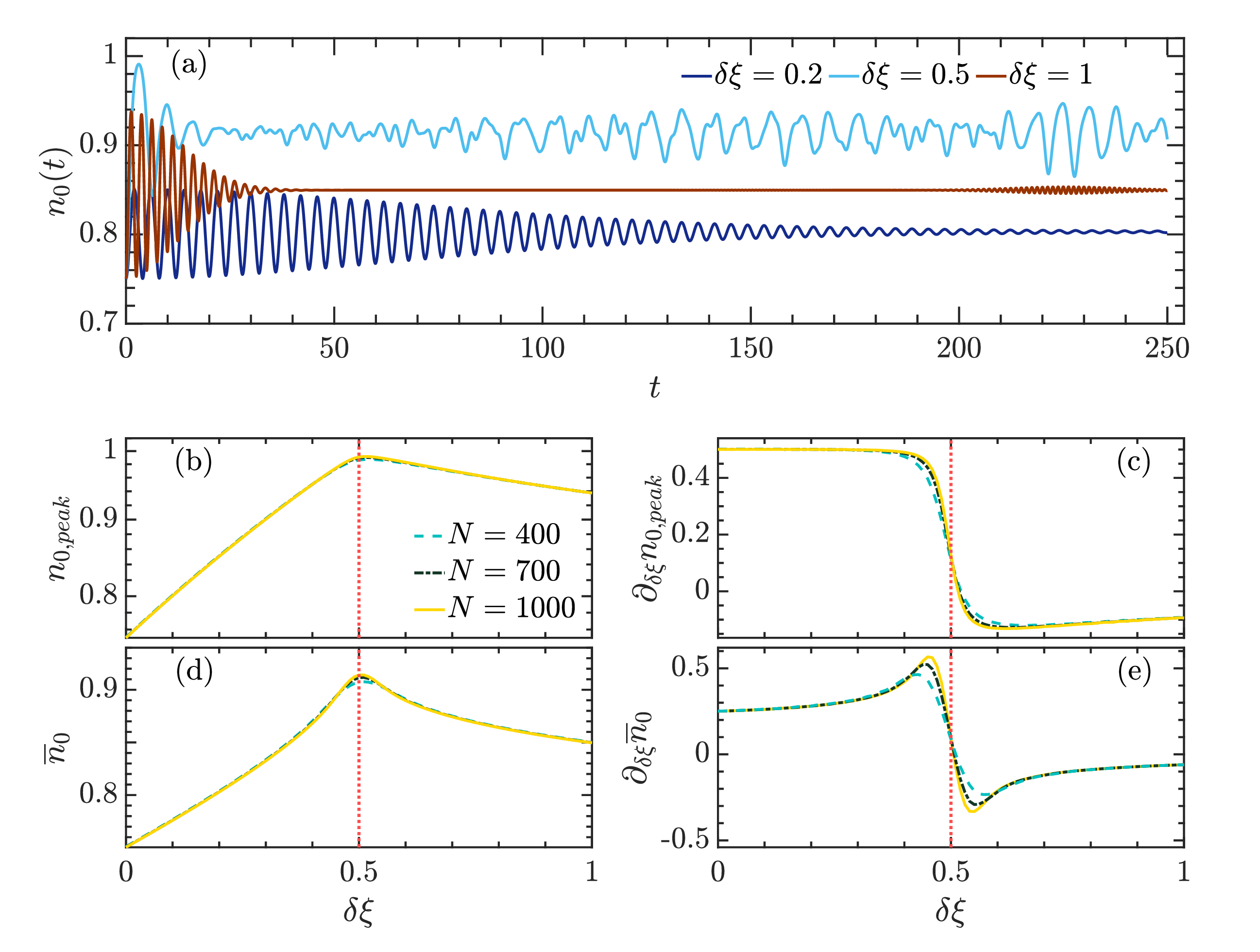}
  \caption{(a) Time evolution of $n_0(t)$ for several values of 
  $\delta\xi$ with $N=1000$.
  (b) First peak of $n_0(t)$, denoted by $n_{0,peak}$, and (c) its derivative with respect to $\delta\xi$,
  denoted by $\partial_{\delta\xi}n_{0,peak}$, as a function of $\delta\xi$ for different system sizes 
  [see the legend in panel (b)]. 
  (d)-(e) Long time average of $n_0(t)$, $\overline{n}_0$, and $\partial_{\delta\xi}\overline{n}_0$
  as a function of $\delta\xi$ for the same system sizes as in panels (b) and (c).
  Here, $\overline{n}_0$ is obtained by averaging $n_0(t)$ in the time interval $t\in[0,1000]$. 
  Other parameter: $\xi_i=1$.  
  Vertical red dotted lines in panels (b)-(e) denote $\delta\xi_c=0.5$,
  obtained from Eq.~(\ref{CriticalXi}).
  The axes in all figures are dimensionless.}
  \label{Dyn0}
 \end{figure}    

In the semiclassical limit $N\to\infty$, the equations of motion of the spin-$1$ BEC 
is given by Eq.~(\ref{EOM}) in the main text.  
The semiclassical dynamics is obtained by solving those equations with 
the initial condition is given by the ground state of the system,
that is, $(\cos\varphi_i,n_{0,i})=[\pm1,(2+\xi_i)/4]$. 
The resulting semiclassical dynamics in the phase space for 
different quenching strengths are plotted in Fig.~\ref{Cdyt}.
Here, we have $\xi_i=1$, so that $\delta\xi_c=0.5$.
As illustrated in Fig.~\ref{Cdyt}(b), the semiclassical dynamics 
shows an obvious separatrix at the critical quench strength.
For quenches that below the critical value, the system does not 
have enough energy to surpass the saddle point, leading to the semiclassical 
dynamics is locked within two disconnected regions, as seen in Fig.~\ref{Cdyt}(a). 
On the contrary, once $\delta\xi>\delta\xi_c$, the system has sufficient energy 
to move freely between both regions. 
As a consequence, two disjointed regions merge together 
and the semiclassical dynamics can visit the whole phase space, as shown in Fig.~\ref{Cdyt}(c).  
It is worth pointing out that the semiclassical dynamics constitutes the skeleton 
of the time evolution of Husimi function in phase space, as illustrated in Fig.~\ref{Hsftime}.      

\section{Dynamics of the density of spin-$0$ component} \label{Appb}

Further characterizations of DPTs can be
revealed by exploring the dynamics of
the denisty of spin-$0$ component, defined as
\be
   n_0=\frac{a_0^\dag a_0}{N}.
\ee
As a conventional detectable quantity in spinor BECs, $n_0$ has been employed 
to experimentally probe the DPT-I 
in antiferromagnetic spin-1 BEC \cite{TianT2020,YangHX2019}. 
Here, we use it to scrutinize the signatures of two types of DPTs in ferromagnetic spin-1 BECs.

 \begin{figure}
  \includegraphics[width=\columnwidth]{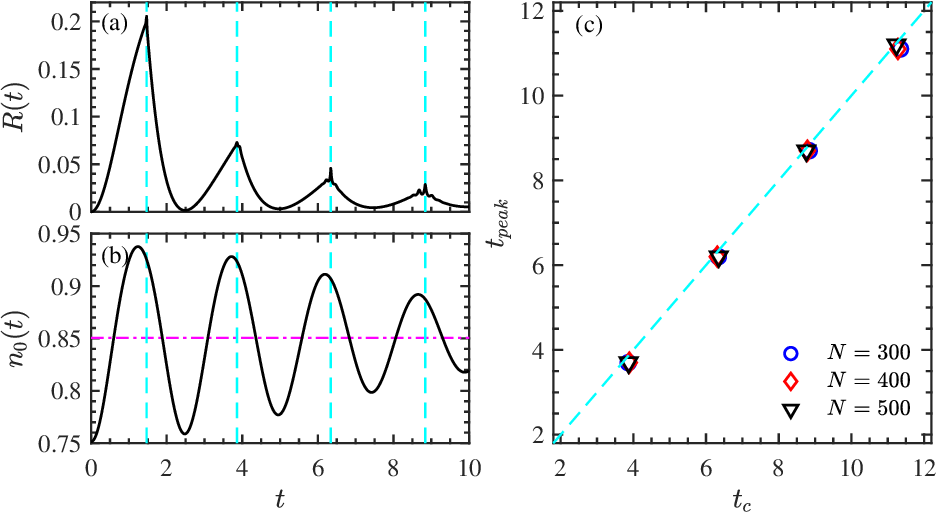}
  \caption{(a) Rate function $R(t)$ and (b) the time evolution $n_0(t)$ 
  for $\delta\xi=1$ with system size $N=300$.
  Vertical dashed lines in panels (a) and (b) marke the critical times of DPT-II,
  while the horizontal dot-dashed line in panel (b) denotes 
  the value of $\overline{n}_0$ in (\ref{DEn0}).
  (c) Times $t_{peak}$, defined as the instants when the local peak in $n_0(t)$ present,
  and the critical times $t_c$ of DPT-II for different system sizes $N$ with $\delta\xi=1$.
  The dashed line corresponds to the function $t_{peak}=t_c$. 
  Other parameter: $\xi_i=1$, associated with $\delta\xi_c=0.5$ [cf.~Eq.~(\ref{CriticalXi})].
  The axes in all figures are dimensionless.}
  \label{nzerodyDPT2}
 \end{figure}

After the quench, the evoultion of $n_0$ is given by
\be \label{Defn0}
   n_0(t)=\mathrm{Tr}[\rho(t)n_0]
      =\sum_{k,m}e^{i(E_k^f-E_m^f)}c_k^\ast c_m n_{0,km},
\ee
where $n_{0,km}=\la E_k^f|n_0|E_m^f\ra$ and $c_s=\la E_s^f|\psi_0\ra$ with $s=m,k$.
The time evolution of $n_0(t)$ for several quenching strengths with $N=1000$ and $\xi_i=1$
is plotted in Fig.~\ref{Dyn0}(a).
One can see that the evolution of $n_0(t)$ shows a remarkable 
change as the value of $\delta\xi$ passes through the critical value of DPT-I, 
which gives by $\delta\xi_c=0.5$ for $\xi_i=1$ case, 
such as we have observed in the dynamics of $M(t)$.
For both $\delta<\delta\xi_c$ and $\delta\xi>\delta\xi_c$ cases, $n_0(t)$ exhibits a regular behavior,
while it irregulary oscillates around a saturation value with 
small amplitude for the critical quenching strength.   

The above observed behaviors of $n_0(t)$ lead us to explore the signatures of DPT-I in the short
and long time evolutions of $n_0(t)$. 
For the short  time case, we focus on the first peak of $n_0(t)$, defined as
$n_{0,peak}=n_0(t=t_{peak})$ with $t_{peak}$ being the time when the first peak present.  
The results shown in Fig.~\ref{Dyn0}(a) indicate that $n_{0,peak}$ should 
have a maximal value at $\delta\xi_c$.
This is indeed what we see in Fig.~\ref{Dyn0}(b), where we plot $n_{0,peak}$
as a function of $\delta\xi$ for different system sizes. 
The presence of DPT-I at $\delta\xi=\delta\xi_c$ can be more clearly revealed by the behavior 
of the derivative of $n_{0,peak}$ with respect to $\delta\xi$.
In Fig.~\ref{Dyn0}(c), we plot 
$\partial_{\delta\xi}n_{0,peak}=\partial n_{0,peak}/\partial(\delta\xi)$ 
as a function of $\delta\xi$ for the same system sizes as in Fig.~\ref{Dyn0}(b).
We see that $\partial_{\delta\xi}n_{0,peak}$ undegoes an obvious jump 
from $0.5$ to a negative value near $\delta\xi_c$, regardless of the system size $N$.
However, the sharpness of the jump increases with increasing $N$. 
Thus, the jump in $\partial_{\delta\xi}n_{0,peak}$ acts as a 
precursor of DPT-I in a finite system. 
It is worth pointing out that $\partial_{\delta\xi}n_{0,peak}$ curves 
for different system sizes cross at $\delta\xi=0.5$, which is in consistence with the
critical point of DPT-I [cf.~Eq.~(\ref{CriticalXi})].

To see how the occurrence of DPT-I gets reflected in the long time dynamics of $n_0(t)$,
we consider the long time average of $n_0(t)$, defined as
\be
  \overline{n}_0=\lim_{T\to\infty}\frac{1}{T}\int_0^Tdtn_0(t).
\ee
Inserting $n_0(t)$ in (\ref{Defn0}) into above equation and employing the fact that 
$E_k\neq E_m$ for $k\neq m$, one  find that $\overline{n}_0$ can be simplified to
\be  \label{DEn0}
  \overline{n}_0=\sum_k|c_k|^2n_{0,kk}.  
\ee
Figure \ref{Dyn0}(d) illustrates how the $\overline{n}_0$ depends on $\delta\xi$
for different system sizes. 
Clearly, the underlying DPT-I results in a peak in the behavior of $\overline{n}_0$, 
suggesting $\overline{n}_0$ can be used as a probe of DPT-I. 
The ability of $\overline{n}_0$ to detecte DPT-I is further verfied by investigating
its derivative with respect to $\delta\xi$.
In Fig.~\ref{Dyn0}(e), we demonstrate how the
$\partial_{\delta\xi}\overline{n}_0$ varies as a function of $\delta\xi$ for several system sizes.
The drmatic change in the behavior of $\partial_{\delta\xi}\overline{n}_0$ is 
clearly visible near the crtical quenching strength.
In particular, the variation of $\partial_{\delta\xi}\overline{n}_0$ with $\delta\xi$ for 
different $N$ also cross at $\delta\xi=0.5$, 
indicating the presence of DPT-I in the thermodynamic limit.
This is in agreement with the analytical result given by Eq.~(\ref{CriticalXi}).

Let us finally discuss the relationship between the dynamics of $n_0$ and the second kind of DPTs.
To this end, we plot the rate function $R(t)$ and $n_0(t)$ in Figs.~\ref{nzerodyDPT2}(a) and
\ref{nzerodyDPT2}(b), respectively, for $\delta\xi=1$ and $N=300$ with $\xi_i=1$.
One can clearly see that the critical times $t_c$ of DPT-II are associated 
with the instants of time $t_{peak}$, corresponding to the location of peaks in $n_0(t)$.
Although the finite system size leads to an obvious deviation between $t_c$ and $t_{peak}$,
the agreement between them should be enhanced by increasing $N$.
We compare the local peak times $t_{peak}$ of $n_0(t)$ to the critical times $t_c$ 
for different system sizes in Fig.~\ref{nzerodyDPT2}(c). 
As shown in the figure, the variation of $t_{peak}$ with $t_c$ 
is well captured by a linear function, which
converges to the form $t_{peak}=t_c$ with increasing $N$.
Hence, the local peak in $n_0(t)$ behaves as a witness of DPT-II.


\bibliographystyle{apsrev4-1}
\bibliography{DQPTBEC}

\end{document}